\documentclass[twocolumn]{article}

\usepackage{arxiv}
\usepackage[utf8]{inputenc} % allow utf-8 input
\usepackage[T1]{fontenc}    % use 8-bit T1 fonts
\usepackage{hyperref}       % hyperlinks
\usepackage{url}            % simple URL typesetting
\usepackage{booktabs}       % professional-quality tables
\usepackage{amsfonts}       % blackboard math symbols
\usepackage{nicefrac}       % compact symbols for 1/2, etc.
\usepackage{microtype}      % microtypography
\usepackage{amssymb}
\usepackage{amsmath}
\usepackage{graphicx}
\usepackage[numbers]{natbib} 

\usepackage{color}

\usepackage{enumerate}
\usepackage{bm}
\usepackage{graphics,epsf,psfrag,pifont,dsfont,bbold}
\usepackage{siunitx}
\usepackage{color}
\usepackage{xspace}
\usepackage{accents}
\usepackage[normalem]{ulem}
\usepackage{float}
\usepackage{balance}
\usepackage{mathptmx}
\usepackage{lastpage}
\usepackage{fancyhdr}
\usepackage{fnpos}
\usepackage{array}
\usepackage{droidsans}
\usepackage{charter}
\usepackage[usenames,dvipsnames]{xcolor}
\usepackage{setspace}
\usepackage[compact]{titlesec}
\usepackage{euscript}

\title{\Large{Rayleigh-B\'{e}nard convection of a model emulsion: anomalous heat-flux fluctuations and finite-size droplets effects}}

\author{
  Francesca Pelusi\thanks{Current affiliation: Forschungszentrum Jülich GmbH, Helmholtz Institute Erlangen-Nürnberg for Renewable Energy (IEK-11) - Cauerstra\ss e 1, 91052 Erlangen, Germany}\\
  Department of Physics \& INFN, University of Rome ``Tor Vergata''\\
  Via della Ricerca Scientifica 1, 00133 Rome, Italy \\
\texttt{f.pelusi@fz-juelich.de}
  \AND
   Mauro Sbragaglia\\
  Department of Physics \& INFN, University of Rome ``Tor Vergata''\\
  Via della Ricerca Scientifica 1, 00133 Rome, Italy \\
  \AND
 Roberto Benzi\\
  Department of Physics \& INFN, University of Rome ``Tor Vergata''\\
  Via della Ricerca Scientifica 1, 00133 Rome, Italy \\
  \AND
  Andrea Scagliarini\\
 Istituto per le Applicazioni del Calcolo, CNR\\
  Via dei Taurini 19, 00185 Rome, Italy\\
  \AND
  Massimo Bernaschi\\
 Istituto per le Applicazioni del Calcolo, CNR\\
  Via dei Taurini 19, 00185 Rome, Italy\\
  \And
  Sauro Succi\\
 Istituto per le Applicazioni del Calcolo, CNR\\
  Via dei Taurini 19, 00185 Rome, Italy\\
  Center for Life Nano Science Sapienza, Istituto Italiano di Tecnologia \\
  Viale Regina Elena 295, 00161 Rome, Italy\\
}
\begin{document}

\twocolumn[
  \begin{@twocolumnfalse}
    \maketitle
    \newpage
    \begin{abstract}
      We present mesoscale numerical simulations of Rayleigh-B\'{e}nard (RB) convection in a two-dimensional model emulsion. The systems under study are constituted of {\it finite-size} droplets, whose concentration $\Phi_0$ is systematically varied from small (Newtonian emulsions) to large values (non-Newtonian emulsions). We focus on the characterisation of the heat transfer properties close to the transition from conductive to convective states, where it is well known that a homogeneous Newtonian system exhibits a  steady flow and a time-independent heat flux. In marked contrast, emulsions exhibit a non-steady dynamics with fluctuations in the heat flux. In this paper, we aim at the characterisation of such non-steady dynamics via detailed studies on the {\it time-averaged heat flux} and its {\it fluctuations}. To quantitatively understand the time-averaged heat flux, we propose a side-by-side comparison between the emulsion system and a single-phase (SP) system, whose viscosity is suitably constructed from the shear rheology of the emulsion. We show that such local closure works well only when a suitable degree of {\it coarse-graining} (at the droplet scale) is introduced in the local viscosity. To delve deeper into the fluctuations in the heat flux, we furthermore propose a side-by-side comparison between a Newtonian emulsion (i.e., with a small droplet concentration) and a non-Newtonian emulsion (i.e., with a large droplet concentration), at fixed time-averaged heat flux. This comparison elucidates that finite-size droplets and the non-Newtonian rheology cooperate to trigger {\it enhanced} heat-flux fluctuations at the droplet scales. These enhanced fluctuations are rooted in the emergence of space correlations among distant droplets, which we highlight via direct measurements of the droplets displacement and the characterisation of the associated correlation function. The observed findings offer insights on heat transfer properties for confined systems possessing finite-size constituents.

    \end{abstract}
  \end{@twocolumnfalse}
]

%\linenumbers
%\tableofcontents
%

\section{\label{sec:intro}Introduction}
Heat transfer in heterogeneous media made of dispersions of one phase (solid, liquid or gaseous) in another liquid phase is of paramount importance for a broad variety of contexts, ranging from everyday life to technological applications~\cite{Hetsroni82,Glicksman94,Chang99,Royon2000,Egolf05,Wang07,Coquard09,Kamath13,Mcclements15,Shao15,Ali19,Wang2019}. Depending on the composition of the dispersed phase, different types of systems can be considered: dispersions of gas bubbles in a continuous liquid phase (e.g., foams or bubbly flows)~\cite{Cohen05,CantatFOAMS,Cohen13,Wang16,Mathai20}, dispersions of droplets in a liquid matrix (e.g.,  emulsions)~\cite{Larson,Walstra98,Gallegos99,Coussot05,Windhab05}, suspensions of particles dispersed in a liquid solvent~\cite{Stickel05,Wagner09,Morris09,Picano14}. The focus of this paper is on the characterisation of the heat transfer properties in a model emulsion, consisting of deformable droplets of a liquid phase dispersed in another continuum phase with the same viscosity. The dynamical behaviour differs at changing the rheological response of the emulsion, the latter being encoded in the flow-curve of the material reporting the stress $\Sigma$ as a function of the shear rate $\dot{\gamma}$, from which the effective viscosity $\eta_{\text{eff}}$ is extracted as $\eta_{\text{eff}}(\dot{\gamma})=d \Sigma(\dot{\gamma})/d \dot{\gamma}$. The rheology, in turn, depends on the droplet concentration: dilute emulsions behave as Newtonian fluids (i.e. $\eta_{\text{eff}}$=const) with a viscosity that increases with the droplet concentration~\cite{Barnes94,Pal2000,Derkach09,Tadros13}. For larger concentrations, non-Newtonian effects emerge: the latter appear in the form of shear-thinning rheology, whereby the viscosity increases as the shear-rate decreases. This non-Newtonian behaviour is even more pronounced at larger droplet concentrations, where the emulsions can be categorised as yield stress materials~\cite{Larson,Pal96,Balmforthetal14,Bonn17}, with a diverging viscosity at small values of $\dot{\gamma}$, while exhibiting a finite viscosity at larger values of $\dot{\gamma}$. The mechanical response of emulsions has been vastly characterised in experiments, theory and simulations, both in the case of Newtonian emulsions as well as in the case of non-Newtonian emulsions, as briefly reviewed below. \\

Regarding Newtonian emulsions, literature offers a very detailed characterisation of the response of the medium (see  ~\cite{Grace82,Rallison84,Stone94,Fischer07,VanPuyvelde08,Minale10,Guido11} for reviews on the topic). One may refer, for example, to the vast knowledge on the deformation and break-up properties of single constituents (i.e., an emulsion in the extremely dilute limit) and/or the characterisation of the medium effective viscosity from dilute to semi-dilute concentrations. However, all these situations typically refer to cases where the material response is analysed in the presence of external drivings, either a force or a shear. As a matter of fact, such a very detailed knowledge is somehow not mirrored in a corresponding characterisation of the heat transfer properties of the medium. A practical case in point is the thermal convection~\cite{Busse78,Grossmann99,Grossmann01,Lappa09}, that we consider in this paper in the widely studied Rayleigh-B\'enard (RB) set-up~\cite{Rayleigh1916,Moore73,Bodenschatz2000,Ahlers09,Lohse10}, consisting of a material between two parallel walls at different temperatures (a hot bottom and a cold top wall). In this situation, the material is driven by buoyancy forces which depend on the local temperature field; the temperature field, in turn, is advected by the velocity field that diffuses in space via the viscosity of the material. For homogeneous Newtonian fluids, an infinitesimal stress perturbation can linearly destabilise the conductive state if the advective time, which takes for a thermal perturbation (a ``plume'') to travel from one wall to the other, is smaller than the time that it takes to be smeared out by thermal diffusion. There exists a critical ratio of these two timescales above which steady convection sets in~\cite{Chandrasekhar61}. Convection has been studied in biphasic systems comprising bubbles~\cite{Biferale12,Orestaetal09,Lohse10,Lakkaraju13} but treating the dispersed objects as if they were point-like or in the dilute limit. Actually, we may expect that, especially in highly confined systems and/or in concentrated dispersions, the {\it granularity} of the system will lead to a failure of any attempt of modelling employing continuum equations or with point-like particles.\\ 

Regarding non-Newtonian emulsions, there is a lot of knowledge pertaining to the mechanical response of such systems under the effects of an external driving~\cite{Rallison84,Fischer07,Bonn17}, but very little is known on the convective heat transport properties. Some studies investigated the convective heat transfer of model-systems exhibiting non-Newtonian rheology (similar to that of highly concentrated emulsions), focusing on the role of the yield stress~\cite{Zhang06,BalmforthRust09,Vikhansky09,Vikhansky10,AlbaalbakiKhayat11,Turanetal12,Davailleetal13,Massmeyer13,Kebicheetal14,Balmforthetal14,Hassanetal15,Karimfazli16}. In these studies, it is shown that, when the rheology changes from Newtonian to non-Newtonian, the stability of the base conductive state changes, to the point that for a yield stress material it becomes linearly stable~\cite{Zhang06,BalmforthRust09,Balmforthetal14} and a finite perturbation intensity is required for the onset of convection; this perturbation value increases upon approaching the Newtonian critical point~\cite{Zhang06}. However, all these theoretical/numerical insights predominantly consider the problem of thermal convection in the presence of ``local'' rheology. In other words, it is assumed that the viscosity that enters the momentum equation depends locally on $\dot{\gamma}$. This assumption may be reasonable whenever convection is treated on ``continuum scales'', i.e., at scales much larger than the characteristic size of the constituents of the material. When we move to scales comparable with that of the constituents, it is known that a description based on a local relation between $\Sigma$ and $\dot{\gamma}$ falls short of capturing the relevant physics, and {\it finite-size} effects need to be taken into account to obtain a comprehensive characterisation of the flow~\cite{Pal96,Goyon08,Nicolas13,BenziSbragaglia16,Derzsi17,LulliBenziSbragaglia18}. It was argued that convective transport of non-Newtonian complex fluids might be impacted in a non-trivial way by "rearrangements" of the mesoscopic constituents at small scales, but unfortunately, due to the limited resolution, the available experimental data were not conclusive~\cite{Davailleetal13}.\\

\begin{figure}[t!]
\centering
\includegraphics[width=.98\linewidth]{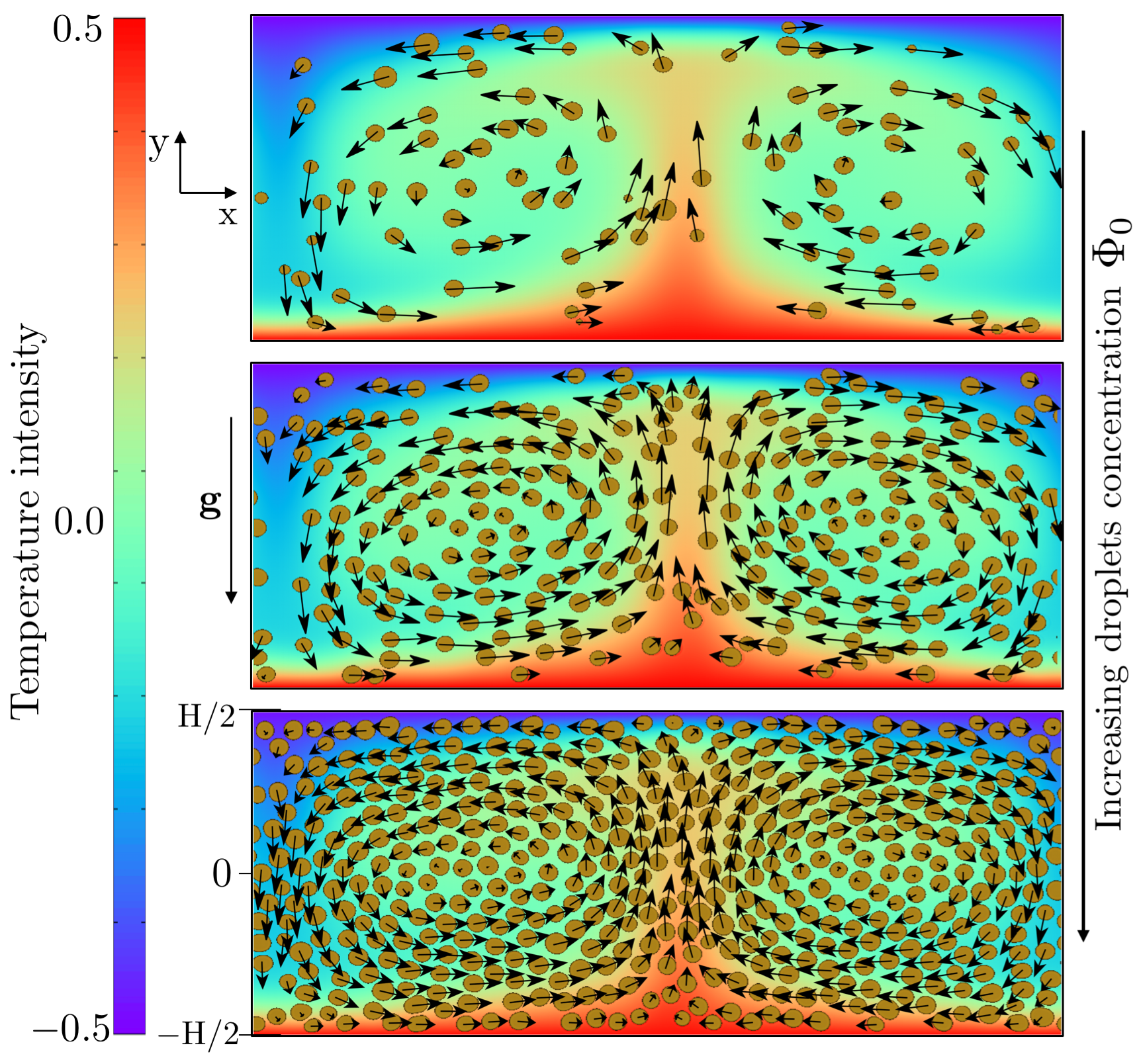}
\caption{Numerical simulations set-up: we study the Rayleigh-B\'{e}nard (RB) convection in two-dimensional concentrated emulsions made of droplets (dispersed phase, dark-yellow domains) into a continuous phase. The emulsions are placed between two parallel walls at fixed temperatures in $y=\pm H/2$, while gravity forces (buoyancy) act along the wall-to-wall direction. We focus on the convective regimes close to the transition from conduction to convection, where the systems display a temperature plume with a two-rolls structure in the velocity field (black arrows represent droplets displacements during convection). We focus on the heat transfer efficiency at fixed buoyancy forces, while changing the concentration $\Phi_0$ of the emulsions, from the dilute to the concentrated regimes. The temperature field is shown in simulation units.}\label{fig:setup}
\end{figure}

The motivation of our study is that both for Newtonian emulsions or non-Newtonian emulsions, there is a lack of knowledge on the characterisation of the convective heat transfer mediated by finite-size droplets. In the RB set-up, such characterisation can be accomplished by considering confined systems, with a wall-to-wall distance $H$ of the order of a few tens of constituents size (cfr. Fig.~\ref{fig:setup}). Close to the transition from conduction to convection, homogeneous Newtonian systems display a steady flow with time-independent heat flux. The emulsions studied in this paper, instead, display a non-stationary heat transfer flux with fluctuations that increase with the concentration. This is accompanied by the development of heterogeneous droplet concentrations across the cell. Intending to characterise and understand quantitatively such heat transfer mechanisms, we studied both the associated {\it time-average} and {\it fluctuations}. Due to the presence of the dispersed phase, emulsions are more viscous than the underlying continuum phase. Hence, it comes naturally to compare these heterogeneous two-phase system with a single-phase (SP) fluid model with some effective viscosity, suitably constructed from the shear rheology of the emulsion. We will show that the heat flux of the SP system can match the time-averaged heat flux of the emulsions only if we introduce a spatial averaging procedure (``coarse-graining'') having a scale of the order of the droplet size. This fact clearly points to the {\it granular} nature of these complex fluids and the necessity to include it in any quantitative characterisation of the heat transfer mechanisms. Regarding the fluctuations, we will study their dependency on the droplet concentrations and also explore fluctuations from "large scales" down to droplets scales. Again, the finite size of the droplets is crucial in promoting the emergence of such fluctuations, which can be remarkably enhanced in conjunction with the non-Newtonian rheology. Our work hinges on numerical simulations that allow an unprecedented detailed analysis of the heat transfer, thereby permitting to highlight both the role of finite-size constituents and their space-time correlations.\\

The paper is organised as follows: in Section~\ref{sec:numerics} we report on the tools for the numerical simulations and the numerical set-ups used (further details can be found in the ESI section); in Section~\ref{sec:rheology} we present a shear-rheology characterisation of the emulsions, that is a necessary pre-requisite to study RB thermal convection; in Section~\ref{sec:heat} we present the phenomenology on the heat transfer in RB convection at changing the droplet concentration; in Section~\ref{sec:results} we quantitatively analyse the time-averaged heat flux and propose an effective modelling for it; in Section~\ref{sec:confronto} we will quantitatively analyse the heat-flux fluctuations; conclusions will be drawn in Section.~\ref{sec:conclusions}.
%
%%%%%%%%%%%%%%%%%%%%%%%%%%%%%%%%%%%%%%%%%%%%%
%%%%%%%%%%%%%%%%%%%%%%%%%%%%%%%%%%%%%%%%%%%%%
%
\section{Methods}\label{sec:numerics}
We report the results of numerical simulations of RB thermal convection in stabilised two-dimensional emulsion systems. The choice of the dimensionality is instrumental to properly resolve the emulsion droplets in the simulation and achieve appropriate statistics in a reasonable amount of time. Moreover, for 2D systems, we can use the Lagrangian tool of analysis developed in~\cite{Bernaschi16}, which is not available presently for three dimensions. A careful inspection of the changes induced by the dimensionality surely requires an additional hard work that warrants future studies. Regarding the numerical technique, we resort to the mesoscale lattice Boltzmann method (LBM)~\cite{Succi18,Kruger17}. Briefly, the model allows the simulation of two-component fluids (say $A$ and $B$, with densities $\rho_A$ and $\rho_B$, respectively) undergoing phase segregation, wherein the system can be divided into bulk regions with the majority of one of the two components. Coalescence of bulk domains is further inhibited by the introduction of repulsive interface forces. Thus, the system can be prepared with a number of droplets $N_{\mbox{\tiny droplets}}$ (dispersed phase) inside the continuous phase (cfr. Fig.~\ref{fig:setup}).\\ 
The concentration of the droplets $\Phi_0$ is a tunable parameter in the preparation of the system, thus we can explore situations ranging from dilute to denser droplet concentrations. $\Phi_0$ is defined as the ratio of the area of the dispersed phase over the total area, $\Phi_0=\mathcal{A}_d/\mathcal{A}_t$; such a definition surmises, of course, a sharp interface. Since we use a diffuse interface method, we need to introduce a threshold, that is $\mathcal{A}_d = \int \int \Theta(\rho_A(x,y)-\rho^*)dx \ dy$, where $\Theta$ is the Heaviside step function and $\rho^*$ is a reference value taken as the mean density of the dispersed ($A$) phase  evaluated between its values in the bulk phases inside and outside the droplets, i.e., $\rho^* = (\rho_{A, \mbox{\tiny bulk}}^{\text{in}}+\rho_{A, \mbox{\tiny bulk}}^{\text{out}})/2$. Further details on the LBM used can be found in the ESI section.\\
Buoyancy forces act on the emulsion. At hydrodynamical scales the reference dynamical equations are the diffuse-interface Navier-Stokes-Boussinesq equation for the hydrodynamical field ${\bf u}(x,y,t)=(u_x,u_y)(x,y,t)$ (repeated indexes are summed upon)
\begin{equation}\label{eq:NS}
\begin{split}
&\rho \left(\partial_t+u_k \partial_k\right) u_i  = \\ &=- \partial_j P_{ij} + \eta_0 \partial_j \left(\partial_{i} u_j+\partial_{j} u_i\right) +\rho \alpha g T \delta_{iy} \qquad i=x,y
\end{split}
\end{equation}
where $\rho$ is the local total density, $P_{ij}$ the non-ideal pressure tensor, $\eta_0$ the dynamic viscosity of the bulk phase~\footnote{This is the viscosity that the system would exhibit in the presence of a homogeneous continuous phase without droplets.}, $\alpha$ the thermal expansion coefficient and $g$ the gravity acceleration.
The temperature field $T(x,y,t)$ (taken as relative to some reference temperature) obeys the advection-diffusion equation 
\begin{equation}\label{eq:T}
\partial_t T + u_k \partial_k T = \kappa \partial_{kk} T.
\end{equation}
where $\kappa$ is the thermal diffusivity. The stabilised emulsions are placed in a confined channel, with the walls in $y=\pm H/2$ and periodic conditions in the $x$-direction. No-slip boundary conditions for the fluid are introduced at the walls, whereas Dirichlet-type boundary conditions are imposed for the temperature fields at the walls, $T(x,y=\pm H/2,t)=\mp \Delta T/2 = \mp 0.5 \ \text{lbu}$ (with $\Delta T = 1.0 \ \text{lbu}$, i.e., lattice Boltzmann units). In  Fig.~\ref{fig:setup} we report some pictorial views of how the system looks like at different concentrations. In all simulations, the Capillary  number ($Ca$) and Reynolds number ($Re$) stay small/moderate ($Ca < 10^{-2}$, $Re < 10^{2}$).\\
The software we employ for all the simulations is an extension of an in-house developed code written in C-CUDA. The code has been described in detail elsewhere \cite{BernaschiGPU09,Bernaschi16}. Here we recall just that it exploits at its best the computing power of modern Graphics Processing Units (GPU) employing an innovative memory access pattern. The code can run on multiple GPUs. To that purpose, we resort to a hybrid parallel programming model (based on a combination of MPI and CUDA). The smoothness by which the thermal LB component has been implemented confirms the flexibility of the software that, besides, supports many different boundary conditions and the chance of simulating the presence of obstacles within the computational domain~\cite{ScagliariniSbragagliaBernaschi15,ourEPL16,PelusiEPL19}.\\ 
Notice that, hereafter, all dimensional observables will be reported in simulation units (i.e., lattice Boltzmann units, lbu).

%%%%%%%%%%%%%%%%%%%%%%%%%%%%%%%%%%%%%%%%%%%%%%%%%%%%%%%%%%%%%%%%%%
\section{Emulsion Rheology}\label{sec:rheology}
%%%%%%%%%%%%%%%%%%%%%%%%%%%%%%%%%%%%%%%%%%%%%%%%%%%%%%%%%%%%%%%%%%

%%%%%%%%%%%%%%%%%%%%%%%%%%%%%%%%%%%%%%%%%%%%%%%%%%%%%%%%%%%%%%%%%%%
\begin{table}
\begin{center}
\begin{tabular}{|p{1.6cm}|p{1.6cm}| |p{1.6cm}|p{1.6cm}|}
\hline
$\Phi_0$ & $N_{\mbox{\tiny droplets}}$  & $\Phi_0$ & $N_{\mbox{\tiny droplets}}$ \\
\hline
0.0735 & 90  & 0.2680 & 284\\
0.1038 & 120 & 0.3322 & 338\\
0.1433 & 159 & 0.3978 & 392\\
0.1721 & 214 & 0.4775 & 449\\
0.2018 & 235 & 0.5413 & 496\\
0.2357 & 242 &  &\\
\hline
\end{tabular}
\caption{Number of droplets $N_{\mbox{\tiny droplets}}$ simulated for each concentration $\Phi_0$.}\label{table}
\end{center}
\end{table}
%%%%%%%%%%%%%%%%%%%%%%%%%%%%%%%%%%%%%%%%%%%%%%%%%%%%%%%%%%%%
%%%%%%%%%%%%%%%%%%%%%%%%%%%%%%%%%%%%%%%%%%%%%
\begin{figure}[t!]
\centering
\begin{tabular}{c}
\includegraphics[width=.97\linewidth]{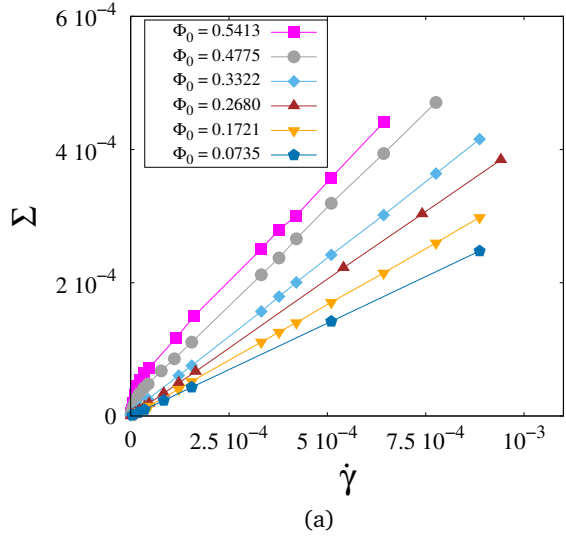}\\
\hspace{0.5cm} \small (a) \\
\includegraphics[width=1.\linewidth]{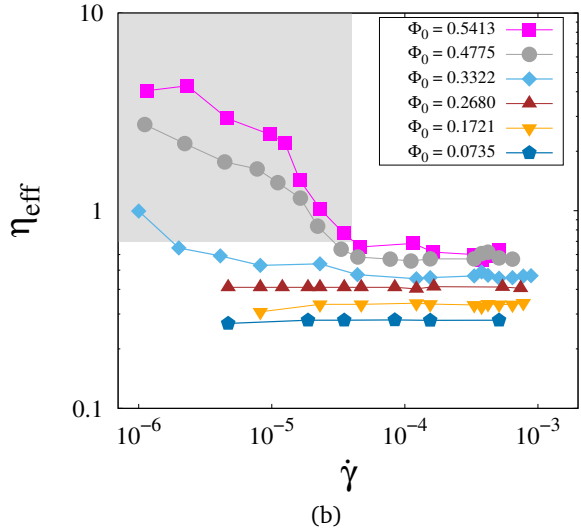}\\
 \hspace{0.5cm} \small (b)
  \end{tabular}
%\hspace{0.8cm}
\caption{Shear rheology of the emulsions. Panel (a): flow curves for the emulsion systems obtained with dedicated shear experiments (see text for details). The concentration is varied. Panel (b): the emulsion effective viscosity $\eta_{\text{eff}}$ as a function of shear rate $\dot{\gamma}$, extracted from the flow curves in panel (a), for different concentrations~\cite{Pal2000}. The dark region refers to a range of concentrations for which non-Newtonian effects are observed. All dimensional quantities are reported in simulation units. \label{fig:shearRheology}}
\end{figure}
%%%%%%%%%%%%%%%%%%%%%%%%%%%%%%%%%%%%%%%%%%%%%
Before performing numerical simulations on convective emulsions at changing the concentration, it is mandatory to perform a rheological characterisation of the systems under study. From one side, this rheological characterisation is useful to compare our data with available literature data (especially in the dilute limit); from the other side, it is also instrumental to provide a characterisation of the functional behaviour of the dynamic viscosity as a function of $\Phi_0$, from dilute to finite concentrations. This will constitute an important point for the study discussed in Sec.~\ref{sec:results}.\\
The rheological characterisation of the emulsions is performed via dedicated experiments in Couette cells, where constant and opposite velocities are imposed at the walls ($u^{\mbox{\tiny wall}}_x(y=\pm H/2,t)=\pm U$. Given the shear rate  $\dot{\gamma}=2 U/H$, we measure the resulting stress $\Sigma$. This allows us to extract the flow curves, i.e., the relation between $\Sigma$ vs. $\dot{\gamma}$. Simulations are performed by placing the emulsions in a channel of height $H \sim 17d$, where $d$ is the mean droplet diameter~\footnote{The mean droplet diameter $d$ is around $24 \ \text{lbu}$.}, and we systematically explore different droplet concentrations $\Phi_0$, from very diluted to concentrated emulsions, by varying the number of droplets $N_{\mbox{\tiny droplets}}$ (see Table~\ref{table}). All the emulsions analysed are pretty monodisperse with tiny variations in the droplet area.
In Fig.~\ref{fig:shearRheology} we show the flow curves (panel (a)), and the effective viscosity (panel (b)) for various concentrations of the emulsions. Given the flow curve data, the effective viscosity is measured as $\eta_{\text{eff}} (\dot{\gamma})=d\Sigma(\dot{\gamma})/d\dot{\gamma}$. 
At low droplet concentrations, the emulsion behaves as a Newtonian fluid ($\eta_{\text{eff}} = \mbox{const}$), with an augmented effective viscosity~\cite{Jeffrey76,Larson}: dark region in Fig.~\ref{fig:shearRheology}(b) (and hereafter) shows the range in which non-Newtonian effects are observed.
%%%%%%%%%%%%%%%%%%%%%%%%%%%%%%%%%%%%%%%%%%%%%
\begin{figure}[t!]
\centering
\includegraphics[width=1.\linewidth]{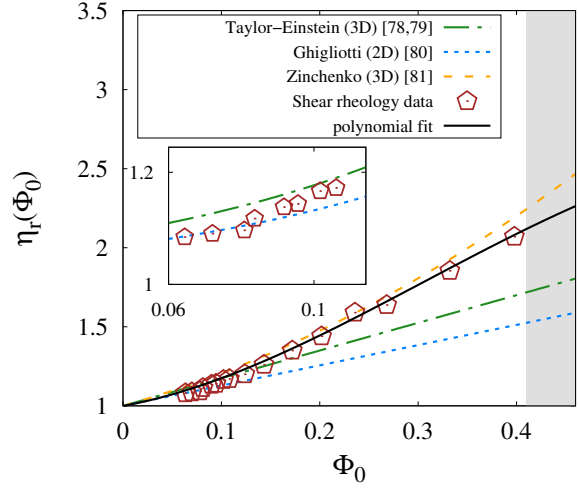}\\
\caption{We report the intrinsic viscosity $\eta_r(\Phi_0)$ (cfr. Eq.~\eqref{eq:TayEin}) for the shear rheology data of panel (a) of Fig.~\ref{fig:shearRheology}, along with its best polynomial fit. We also report literature estimates for the effective viscosity, both in a 2D and in a 3D set-up. In the inset, we zoom-in at very low $\Phi_0$ to highlight the agreement with~\cite{Ghigliottietal10} up to $\Phi_0 \approx 0.12$. The dark region refers to a range of concentrations for which non-Newtonian effects start to emerge (cfr. Fig.~\ref{fig:shearRheology}).\label{fig:initrinsicViscosity}}
\end{figure}
%%%%%%%%%%%%%%%%%%%%%%%%%%%%%%%%%%%%%%%%%%%%%
By focusing on the Newtonian emulsions presenting a linear rheology, in Fig.~\ref{fig:initrinsicViscosity} we compare the effective viscosity extracted from simulations with the literature data. Let us recall that for suspensions of solid spherical particles, in the very dilute limit ($\Phi_0 \rightarrow 0$), the three-dimensional Einstein relation predicted a linear growth of the relative viscosity with $\Phi_0$~\cite{Einstein1906}; later on, G.I. Taylor proved that linearity holds also for the relative viscosity of three-dimensional emulsions~\cite{Taylor32} (in the small droplet deformation regime), i.e.,:
\begin{equation}\label{eq:TayEin}
\eta_r(\Phi_0) \equiv \frac{\eta_{\text{eff}}(\Phi_0)}{\eta_{\text{solv}}} = 1+ [\eta]_0 \Phi_0,
\end{equation}
with an intrinsic viscosity coefficient dependent on the viscosity ratio $\lambda$ as $[\eta]_0 = \frac{\frac{5}{2}\lambda+1}{\lambda + 1}$ (which tends to $5/2$, Einstein's coefficient for solid particles, as $\lambda \rightarrow \infty$). The measured relative viscosity is in good agreement with Eq.~\eqref{eq:TayEin}, with $[\eta]_0 =7/4$, as expected for an emulsion with unitary viscosity ratio ($\lambda = 1$), for concentrations up to $\Phi_0 \approx 0.12$. The agreement is improved upon using a 2D estimate of the effective viscosity, that we have extracted from the data in~\cite{Ghigliottietal10}, as we can see in the zoom-in reported in the inset. At larger droplet concentrations, data start to deviate from dilute predictions. Specifically, for larger $\Phi_0$ (and up to $\Phi_0 \approx 0.35$) our data agree well with Zinchenko's prediction for three-dimensional emulsions~\cite{Zinchenko84}. 

\section{Heat transfer phenomenology}\label{sec:heat}
%%%%%%%%%%%%%%%%%%%%%%%%%%%%%%%%%%%%%%%%%%%%%%%

\begin{figure}[t!]
\centering
\begin{tabular}{c}
\includegraphics[width=1.\linewidth]{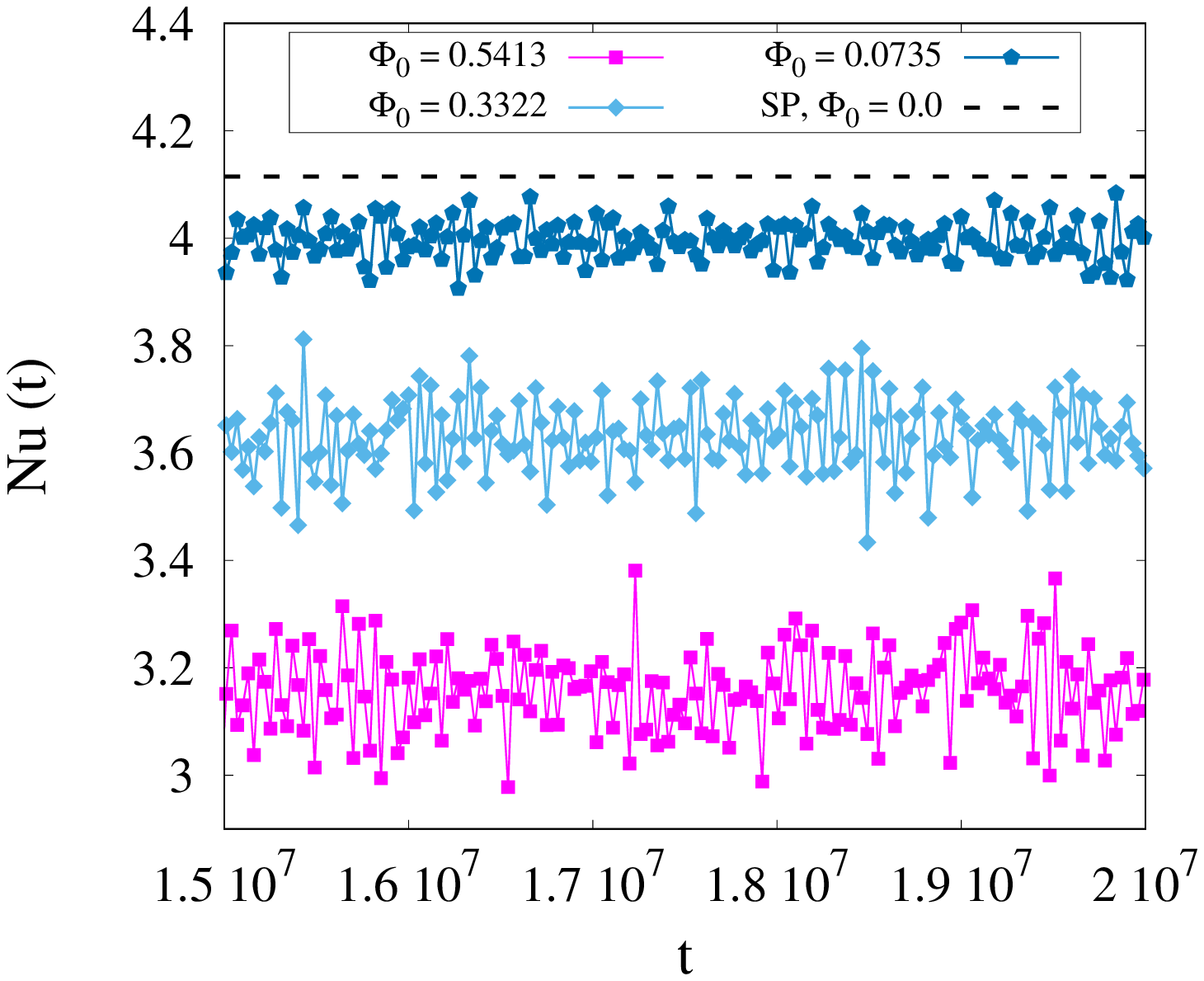}\\
\small (a) \\
\includegraphics[width=1.\linewidth]{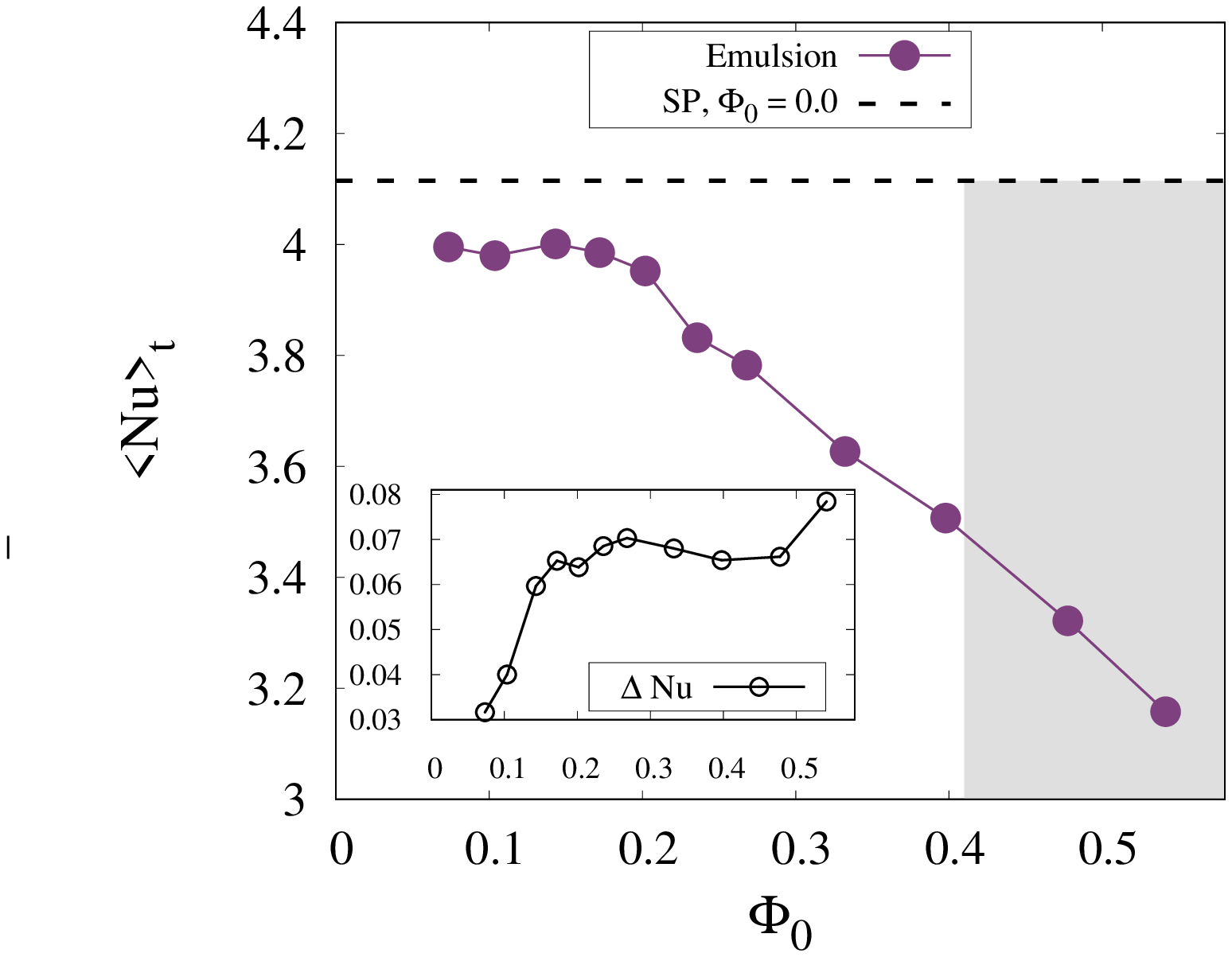}\\
\small (b) \\
\end{tabular}
\caption{Panel (a): Nusselt number as a function of time $t$ (cfr. Eq.~\eqref{eq:NuTime}) (time shown in lbu) for different concentrations $\Phi_0$. The magnitude of buoyancy forces is kept fixed. We report in the $x$-axis a representative time lapse where the system has reached a statistically steady state. The dashed line represents the corresponding Nusselt number for a single-phase (SP) system ($\Phi_0 \rightarrow 0$). Panel (b): we report the time-average of the Nusselt number $\langle \mbox{Nu} \rangle_t$. In the inset, we report the time-averaged fluctuations of the Nusselt number with respect to its time average, $\Delta \mbox{Nu} =  \langle \left (\mbox{Nu}(t) - \langle \mbox{Nu}\rangle_t \right)^2\rangle_t^{1/2}$. The dark region refers to a range of concentrations for which non-Newtonian effects start to emerge (cfr. Fig.~\ref{fig:shearRheology}).\label{fig:NuTime}}
\end{figure}

In this section, we provide an overview of the properties of heat transfer at changing the concentration, from dilute to larger concentrations. In order to assess the heat transfer properties of the emulsion, we focus on the heat flux across the system, $\mathcal{F}$, which is the sum of a conductive and a convective part, $\mathcal{F} = \mathcal{F}_{\text{cond}} + \mathcal{F}_{\text{conv}}$; both can, in principle, differ in the biphasic system, from the mono-phasic counterpart. For our simulations, the two fluids have the same thermal diffusivity and no interfacial thermal resistance is supported, therefore $\mathcal{F}_{\text{cond}}$ is not affected. An obvious effect of increasing the concentration is to increase the system viscosity, as discussed in Sec.~\ref{sec:rheology}. The increase in viscosity will result in a reduction of the emulsion propensity to convection. Given this, a naive expectation would be that of a monotonic decay of the heat flux with $\Phi_0$. Actually, the phenomenology is richer, because of the emergence of temporal fluctuations in the heat transfer properties. This is quantitatively elucidated in Fig.~\ref{fig:NuTime},where we analyse the heat fluxes, expressed in a dimensionless form via the Nusselt number~\cite{Shraiman90,Ahlers09,Verzicco10,Chilla12}
\begin{equation}\label{eq:NuTime}
\mbox{Nu}(t) = \frac{\langle u_y (x,y,t) T (x,y,t)\rangle_{x,y} - \kappa \langle \partial_y T (x,y,t) \rangle_{x,y}}{\kappa \frac{\Delta T}{H}}
\end{equation}
where $\langle (\dots) \rangle_{x,y}$ denotes a space average. $\mbox{Nu}$ is a parameter that quantifies the relative intensity between convective and conductive transport.
%%%%%%%%%%%%%%%%%%%%%%%%%%%%%%%%%%%%%%%%%%%%%
%%%%%%%%%%%%%%%%%%%%%%%%%%%%%%%%%%%%%%%%%%%%%
In Fig.~\ref{fig:NuTime}(a) we report $\mbox{Nu}(t)$ for different values of $\Phi_0$, at fixed buoyancy amplitude $\alpha g \Delta T = 1.86 \ 10^{-5} \; \text{lbu}$; the time average of $\mbox{Nu}(t)$ over the statistically steady-state ($\langle\mbox{Nu}\rangle_t$) is reported in Fig.~\ref{fig:NuTime}(b), while time-averaged fluctuations with respect to its time average ($\Delta \mbox{Nu}= \langle(\mbox{Nu}(t) - \langle\mbox{Nu}\rangle_t)^2\rangle_t^{1/2}$) are displayed in the inset of the panel (b) as a function of $\Phi_0$. 
We observe that $\langle\mbox{Nu}\rangle_t$ stays nearly constant for concentrations up to $\Phi_0 \approx 0.2$ and then decreases when $\Phi_0$ increases, whereas the fluctuations $\Delta \mbox{Nu}$ tend to increase with $\Phi_0$. In particular, in the limit $\Phi_0\rightarrow 0$, i.e., for a single-phase (SP) system, the fluctuations go to zero, indicating that the convection is stationary (dashed black line in Fig.~\ref{fig:NuTime}), because of a relatively low Rayleigh number, $Ra \approx 5.3 \times 10^4$~\cite{Rayleigh1916}~\footnote{The Rayleigh number $Ra$ is defined as $Ra = \frac{g \alpha \Delta T H^{3}}{\kappa \nu}$, where $\nu$ is the kinematic viscosity. It provides information on the balance between buoyancy force and viscous friction force, so it governs the transition from a conductive to a convective state in an homogeneous system~\cite{Chandrasekhar61}.} The emergence of fluctuations must then be interpreted as a genuine feature of the heterogeneous system and it is ascribed to the presence of the droplet phase.
%%%%%%%%%%%%%%%%%%%%%%%%%%%%%%%%%%%%%%%%%%%%%
\begin{figure}[t!]
\centering
\includegraphics[width=1.\linewidth]{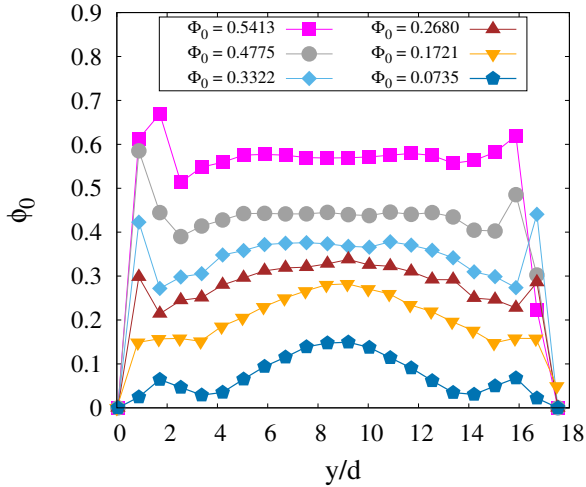}
\caption{Concentration profiles $\phi_0(y)$, obtained by averaging on time and $x$-direction. The $y$ coordinate is normalised to the mean droplet diameter $d$. Each colour/symbol is associated with a different concentration $\Phi_0$.}\label{fig:phiProfiles}
\end{figure}
%%%%%%%%%%%%%%%%%%%%%%%%%%%%%%%%%%%%%%%%%%%%%
Results reported in Fig.~\ref{fig:NuTime} represent a "large scale" characterisation of the heat transfer properties of the studied emulsions, in that they refer to a global observable, i.e. the Nusselt number that is defined as an average over the whole system size (cfr. Eq.~\eqref{eq:NuTime}). Since our system is characterised by finite-size constituents, it comes naturally to inspect properties at smaller scales, comparable to the droplet size. Indeed, it is worth reminding that for the thermal convection in a different soft system (polymer solutions), it was shown that variation in the heat flux could be understood in terms of a space-dependent effective viscosity (due, in that case, to the differential stretching of the polymers along with the cell height)~\cite{BenziChing16}. Inspired by this observation, we monitored the droplet concentrations in the wall-to-wall coordinate ($y$), by averaging over time and along the mainstream flow direction ($x$), during the convection state. The resulting concentration profiles $\phi_0(y)$ are reported in Fig.~\ref{fig:phiProfiles}. We observe, indeed, they are not constant and exhibit a height-varying modulation, especially from low to moderate concentrations $\Phi_0$. Importantly, the modulation with the height of concentration profiles shows a variation on the scale of the droplet, i.e. when the coordinate $y$ changes by an amount of the order of $d$. The development of these non-homogeneous concentration profiles might be due to multiple factors, such as droplet migration induced by a non-uniform shear field~\cite{Hudson03,Jaensson18,Malipeddi19} (owing to the large scale circulation of convection) or droplet depletion due to interactions with the walls. A precise description disentangling these various mechanisms and discriminating which one contributes most lies beyond the scope of the present work. Here, just take the emergence of such profiles as an empirical fact. This said it is nevertheless clear that the non-homogeneity relies on the fact that droplets are transported by the flow. Large $\Phi_0$ implies reduced mobility of the droplets, which is reflected in a relative suppression of profile modulation.\\
Summarising, emulsions with different droplet concentrations exhibit heterogeneous profiles and a time-averaged Nusselt number that decreases (for fixed buoyancy forces) upon increasing the droplet concentration. This decrease in the Nusselt number goes along with increasing fluctuations in the heat flux. In the next sections, we will inspect more quantitatively both the decrease in the time-averaged Nusselt number and the emergence of fluctuations, starting from the "large scale" observations of Fig.~\ref{fig:NuTime} down to smaller scales, comparable to the droplet size.

%%%%%%%%%%%%%%%%%%%%%%%%%%%%%%%%%%%%%%%%%%%%%%%%%%%%%%%%%%%%%%%%%%
\section{Time-Averaged Nusselt Number: effective modelling at the droplet scale}\label{sec:results}
%%%%%%%%%%%%%%%%%%%%%%%%%%%%%%%%%%%%%%%%%%%%%%%%%%%%%%%%%%%%%%%%%%

To delve deeper into the behaviour of $\langle\mbox{Nu}\rangle_t$ with increasing droplet concentration, the natural question we asked is whether one might capture it employing a continuum approach. For this purpose, we ran simulations with the SP system with a homogeneous viscosity equal to the shear viscosity that we have measured (cfr. Fig.~\ref{fig:shearRheology}(b)), i.e.,
\begin{equation}\label{eq:choice1}
\eta_{\text{rheo}}^{\text{SP}}(\Phi_0) = \eta_{\text{eff}}(\Phi_0).
\end{equation} 
This is possible within our numerical approach, by changing the relaxation time of the lattice Boltzmann equation in such a way that the corresponding dynamic viscosity (cfr. Eq.~(13) in ESI section) matches the measured shear viscosity {\it homogeneously} throughout the system. Notice that SP systems constructed in that way exhibit a Nusselt number independent of time, for the reasons posited before. Moreover, we notice that this study is well-posed only for Newtonian emulsions, where the effective viscosity is a function of the concentration only. For non-Newtonian emulsions, one should come up with some refined proposal including the shear rate dependency of the effective viscosity. Therefore, hereafter we will limit the discussion to Newtonian emulsions. Generalisations to Non-Newtonian emulsions will be discussed at the end of this section and in Section~\ref{sec:confronto}.\\
%%%%%%%%%%%%%%%%%%%%%%%%%%%%%%%%%%%%%%%%%%%%%
\begin{figure}[t!]
\centering
\includegraphics[width=.98\linewidth]{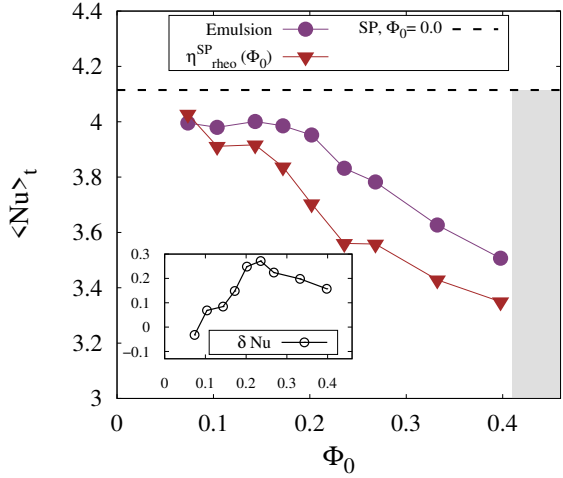}
\caption{The time-averaged Nusselt number $\langle\mbox{Nu}\rangle_t$ as a function of the concentration $\Phi_0$ for two different cases: emulsion (purple circles) and a single-phase system (SP, brown triangles) with a dynamic viscosity that homogeneously matches the rheological viscosity (cfr. Eq.~\eqref{eq:choice1} and Fig.~\ref{fig:shearRheology}). The dark region refers to a range of concentrations for which non-Newtonian effects start to emerge (cfr. Fig.~\ref{fig:shearRheology}). The comparison, exalted by $\delta \mbox{Nu} =\langle \mbox{Nu}\rangle_t-\mbox{Nu}^{\text{\tiny SP}}_{\text{rheo}}$, is given in the inset.}\label{fig:NuPhi0}
\end{figure}
%%%%%%%%%%%%%%%%%%%%%%%%%%%%%%%%%%%%%%%%%%%%%
In Fig.~\ref{fig:NuPhi0} we report the time-averaged Nusselt number $\langle\mbox{Nu}\rangle_t$ as a function of the droplet concentration $\Phi_0$ for both the heterogeneous emulsions and the homogeneous SP system. In the limit $\Phi_0 \rightarrow 0$ the time-averaged Nusselt numbers tend to coincide, as they should. At increasing $\Phi_0$, the Nusselt number measured in the SP simulations decreases monotonically, as expected for an increasingly viscous system. However, we observe a mismatch with the behaviour of $\langle\mbox{Nu}\rangle_t$ in the emulsion case, which becomes particularly evident (with deviations up to roughly 10\%) for intermediate values of the concentration and then decreases again at larger $\Phi_0$.
The assumption of a global effective viscosity equal to the one extracted from the shear rheology is clearly not enough. We then inspected whether a {\it local} effective viscosity should be considered in our case as well, due to a non-homogeneous droplet concentration distribution across the system, as shown in Fig.~\ref{fig:phiProfiles}.
To account for this aspect in the SP fluid model, we promote the effective viscosity to be a local quantity as well, i.e.,
\begin{equation}\label{eq:choice2}
\eta_0^{\text{SP}}(y) = f(\phi_0(y)),
\end{equation}
where the function corresponds to the fit displayed in Fig.~\ref{fig:initrinsicViscosity} (black solid line). By analysing the Nusselt number obtained from SP simulations with the prescription~\eqref{eq:choice2}, we observed that the emulsions data stay in between the two protocols (see Fig. 1 in the ESI section): while the protocol~\eqref{eq:choice1} underestimates the emulsions data, the new protocol~\eqref{eq:choice2} overestimates them. A possible explanation for the deviations observed in comparing the emulsions data with the protocol~\eqref{eq:choice2} can be grasped by looking again at the concentration profiles in Fig.~\ref{fig:phiProfiles}: the latter displays a bulk profile with an overshooting occurring in the wall-proximal regions whose extension is comparable to $1-2$ droplet sizes (droplet layering). This means that forcing the continuum model SP fluid to vary its effective viscosity over such small scales may have a non-trivial effect on the system dynamics.
\begin{figure}[t!]
\centering
\includegraphics[width=.99\linewidth]{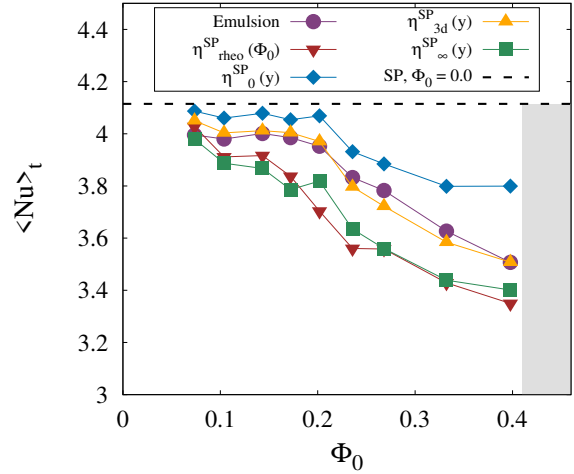}
\caption{The time-averaged Nusselt number $\langle \mbox{Nu}\rangle_t$ as a function of concentration $\Phi_0$ for emulsion (purple circles) and a single-phase system (SP) with different choices of the viscosity $\eta^{\text{\tiny SP}}$ (other colours/symbols). We report data based on $\eta_{\text{rheo}}^{\text{SP}}(\Phi_0)$ from protocol Eq.~\eqref{eq:choice1} and data based on $\eta_{\Lambda}^{\mbox{\tiny SP}}(y)$ from protocol Eq.~\eqref{eq:choice3} using various resolutions of the coarse-graining parameter $\Lambda$ (cfr. Eq.~\eqref{eq:coarsegrain}). The dark region refers to a range of concentrations for which non-Newtonian effects start to emerge (cfr. Fig.~\ref{fig:shearRheology}). } \label{fig:NuPhi}
\end{figure}
%%%%%%%%%%%%%%%%%%%%%%%%%%%%%%%%%%%%%%%%%%%%%
%%%%%%%%%%%%%%%%%%%%%%%%%%%%%%%%%%%%%%%%%%%%%
\begin{figure}[t!]
\centering
\includegraphics[width=.99\linewidth]{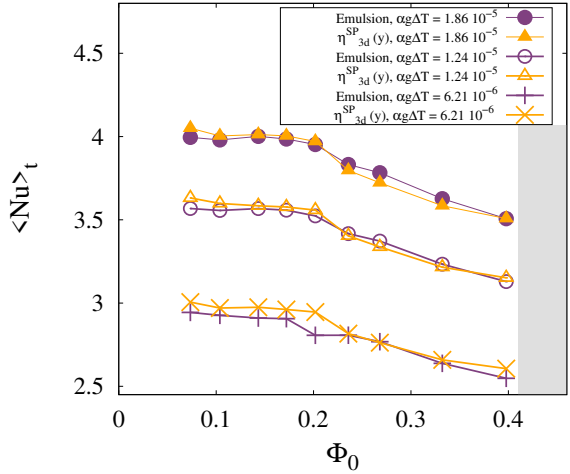}
\caption{The time-averaged Nusselt number $\langle \mbox{Nu}\rangle_t$ as a function of the concentration $\Phi_0$: comparison between emulsion (purple points: full circles, empty circles, daggers) and single-phase system (SP, orange points: full triangles, empty triangles, times symbols) with a resolution of coarse-graining process $\Lambda = 3d$, for different applied buoyancy forces $\alpha g$ in Eq.~\eqref{eq:NS}. The buoyancy amplitudes are reported in simulation units. The dark region refers to a range of concentrations for which non-Newtonian effects start to emerge (cfr. Fig.~\ref{fig:shearRheology}).} \label{fig:diffAlphaG}
\end{figure}
%%%%%%%%%%%%%%%%%%%%%%%%%%%%%%%%%%%%%%%%%%%%%
%%%%%%%%%%%%%%%%%%%%%%%%%%%%%%%%%%%%%%%%%%%%%
\begin{figure*}[t!]
\begin{center}
%\begin{minipage}{1.1\textwidth}
\includegraphics[width=.93\linewidth]{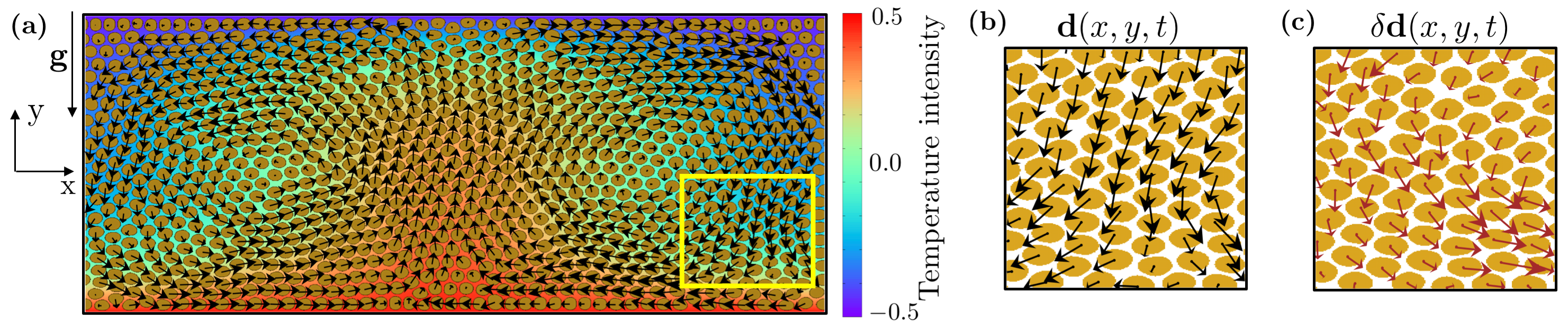}\\
\vspace{0.2cm}
%\begin{minipage}{18cm}
\caption{Panel (a): pictorial view of a non-Newtonian emulsion (NNE, $\Phi_0 = 0.6373$) with the vectorial Lagrangian droplets displacements (black arrows). Starting from the Eulerian droplets displacements ${\bf d}(x,y,t)$ (see panel (b) for a zoom), we construct the associated fluctuations with respect to its time average $\delta {\bf d}(x,y,t)={\bf d}(x,y,t)-\langle {\bf d}(x,y,t) \rangle_t$ (panel (c), see text for more details). All dimensional quantities are reported in simulation units.}\label{fig:setup2}
%\end{minipage}
%\end{minipage}
\end{center}
\end{figure*}
%%%%%%%%%%%%%%%%%%%%%%%%%%%%%%%%%%%%%%%%%%%%%
To overcome this problem, we propose to generalise the local effective viscosity~\eqref{eq:choice2} as follows:
\begin{equation}\label{eq:choice3}
\eta_{\Lambda}^{\text{SP}}(y) = f(\phi_{\Lambda}(y)),
\end{equation}  
where $\phi_{\Lambda}(y)$ is a {\it coarse-grained} (over a size $\Lambda$) concentration profile defined as:
\begin{equation}\label{eq:coarsegrain}
\phi_{\Lambda}(y) = \frac{1}{\Lambda}\int_{y-\Lambda/2}^{y+\Lambda/2} \phi_0(y')dy'.
\end{equation}  
Notice that for $\Lambda = 0$ the original concentration profile $\phi_0(y)$ is recovered, by definition. Evidence of oscillations of $\eta_{\Lambda}^{\text{SP}}(y) = f(\phi_{\Lambda}(y))$ near the walls, stemming from the droplet layering, can be seen in Fig. 2 in the ESI section. These oscillations are smoothed out by the coarse-graining procedure, highlighting that the relative variation of the effective viscosity is actually more important for the lower concentrations (Fig. 2 in the ESI section). Simulating the SP fluid with the choice~\eqref{eq:choice3} for the effective viscosity, indeed, yields the best agreement with the phenomenology of the emulsion in terms of the time-averaged heat flux $\langle \mbox{Nu}\rangle_t$, as shown in  Fig.~\ref{fig:NuPhi}, for $\Lambda = 3d$ (we also plot the data for $\Lambda \rightarrow \infty$ which, not surprisingly, basically overlap with those for $\eta_{\text{rheo}}^{\text{SP}}(\Phi_0)$).\\
As mentioned above, the results obtained so far refer to a fixed buoyancy amplitude $\alpha g \Delta T = 1.86 \ 10^{-5} \; \text{lbu}$ (i.e., at fixed Rayleigh number $Ra$).
It appears then natural to investigate the impact of changing the value of $\alpha g \Delta T$ on the protocol~\eqref{eq:choice3}. To this aim, we have performed additional numerical simulations at different buoyancy amplitude $\alpha g \Delta T$ and compared the time-averaged Nusselt number $\langle\mbox{Nu}\rangle_t$ obtained from simulations of the emulsions at changing $\Phi_0$, with that of SP system with viscosity given by~\eqref{eq:choice3}. Results are reported in Fig.~\ref{fig:diffAlphaG}, which displays a satisfactory agreement between the numerical simulations and protocol~\eqref{eq:choice3} for $\alpha g \Delta T$ spanning roughly an order of magnitude, from $\alpha g \Delta T = 6.21 \ 10^{-6} \; \text{lbu}$ to $\alpha g \Delta T = 1.86 \ 10^{-5} \; \text{lbu}$. Because of the coalescence of the droplets, larger values of the buoyancy amplitude $\alpha g \Delta T$ could not be explored in detail.\\
Before closing this section, we stress that the protocol~\eqref{eq:choice3} has been studied in the framework of Newtonian emulsions, where the effective viscosity $\eta_{\text{eff}}(\Phi_0)$ is independent of the shear rate (cfr. Fig.~\ref{fig:shearRheology}). For non-Newtonian systems, it is necessary to consider the extra complication of a shear-dependent viscosity (see also Section~\ref{sec:confronto} and Fig. 3 in the ESI section). All the attempts that we made in this direction failed in reproducing the time-averaged Nusselt number: some quantitative indications on this point will be shown in the next section. We will also return on this issue in the conclusions.
%%%%%%%%%%%%%%%%%%%%%%%%%%%%%%%%%%%%%%
\section{Anomalous heat transfer fluctuations: from large scales to the droplet scale}\label{sec:confronto}
%%%%%%%%%%%%%%%%%%%%%%%%%%%%%%%%%%%%%%% 
In this section, we study heat transfer fluctuations highlighting the difference between the cases of low and high concentration of droplets. To this aim, we decided to proceed with a side-by-side comparison between a diluted Newtonian emulsion (NE, hereafter) with $\Phi_0 = 0.2680$ and a concentrated non-Newtonian emulsion (NNE, hereafter). The droplet concentration of the NNE is the largest one for which we do not observe coalescence events during the time dynamics~\footnote{For the sake of a fair comparison between NNE and NE, we decided not to consider larger concentrations.} Moreover, in order to maximise the $\Phi_0$ for the NNE, we found that a slightly wider confinement  $H/d \sim 25$ allowed to increase the droplet concentration while retaining the possibility to simulate stable convective states. The chosen $\Phi_0$ for the NNE is $\Phi_0 = 0.6373$, see Fig.~\ref{fig:setup2}(a) to get a pictorial view of how the highly concentrated system looks like. In Fig. 3 in the ESI section we report the flow curves from shear rheology measurements on the two types of emulsion. Notice that the droplet concentration for the NNE is large enough to detect "incipient" yield stress behaviour~\cite{Bonn17}. The buoyancy amplitude, $\alpha g \Delta T$, is chosen in such a way that the system sustains a convective state, close to the transition from conduction to convection. For the analysis we are going to present, it has been necessary to track droplets trajectories and analyse the correlation functions of their displacements. This is possible in our code since it is equipped with a Lagrangian tool of analysis, which allows keeping track of the vectorial displacement of all droplets ${\bf d}_i(t)$ ($i=1...N_{\mbox{\tiny droplets}}$) at all times (cfr. Fig.~\ref{fig:setup2}(a)). From the Lagrangian droplets displacement, we construct the corresponding Eulerian quantity ${\bf d}(x,y,t)$ by considering - for each point $(x,y)$ at a given time $t$ - all droplets displacements that are near that point at that time (cfr. Fig.~\ref{fig:setup2}(b)).
%%%%%%%%%%%%%%%%%%%%%%%%%%%%%%%%%%%%%%%%%%%%%%%%%%%%%%%%%%
\begin{table}[t!]
\begin{center}
\begin{tabular}{|p{3.5cm}|p{1.5cm}|p{2.cm}|}
%\begin{center}
\hline 
Type & $\langle \mbox{Nu}\rangle_t$ &  $\alpha g \Delta T$ \\
\hline
NNE ($\Phi_0 = 0.6373$) & 2.0 & $5.96 \ 10^{-6}$ \\
NE ($\Phi_0 = 0.2680$) & 2.0 & $6.65 \ 10^{-7}$ \\
NNE ($\Phi_0 = 0.6373$) &  2.7 &  $8.71 \ 10^{-6}$ \\
NE ($\Phi_0 = 0.2680$) &  2.7 &  $1.42 \ 10^{-6}$ \\
$\mbox{SP}_{\mbox{\tiny loc}}$ &  2.7 & $3.51 \ 10^{-6}$ \\
\hline
%\end{center}
\end{tabular}
\caption{We report the buoyancy amplitudes necessary to obtain the same time-averaged Nusselt number $\langle \mbox{Nu} \rangle_t$ (see text for more details). The buoyancy amplitudes are reported in simulation units.}\label{table2}
\end{center}
\end{table}
%%%%%%%%%%%%%%%%%%%%%%%%%%%%%%%%%%%%%%%%%
The vectorial displacement ${\bf d}(x,y,t)$ can be averaged in time ($\langle {\bf d}(x,y,t) \rangle_t$) and fluctuations with respect to this time-averaged can be studied (cfr. Fig.~\ref{fig:setup2}(c))
\begin{equation}\label{eq:displacementU}
\delta {\bf d}(x,y,t)={\bf d}(x,y,t)-\langle {\bf d}(x,y,t) \rangle_t.
\end{equation}
In order to address the heat transfer properties, we focus again on the time-dependent Nusselt number (cfr. Eq.~\eqref{eq:NuTime}). Our initial strategy was to compare the fluctuations in the two emulsions at the same heat transport efficiency (i.e., same Nusselt number).  The two emulsions have different effective viscosity, with NNE being more viscous that NE, hence they respond differently to a given imposed buoyancy amplitude. Therefore, we have to determine -- for each emulsion-- the buoyancy amplitude necessary to obtain the desired value of the time-averaged Nusselt number $\langle \mbox{Nu} \rangle_t$ (see Table~\ref{table2} for details). Specifically, it is necessary to impose a larger buoyancy in the dynamical evolution (cfr. Eq.~\eqref{eq:NS}) if the emulsion is more concentrated.
%%%%%%%%%%%%%%%%%%%%%%%%%%%%%%%%%%%%%%%%%%%%%%%%%%%%%%%%%%%%
\begin{figure*}[t!]
\centering
%\begin{minipage}[t]{1.\textwidth}
\begin{tabular}{c c c}
\includegraphics[width=.315\linewidth]{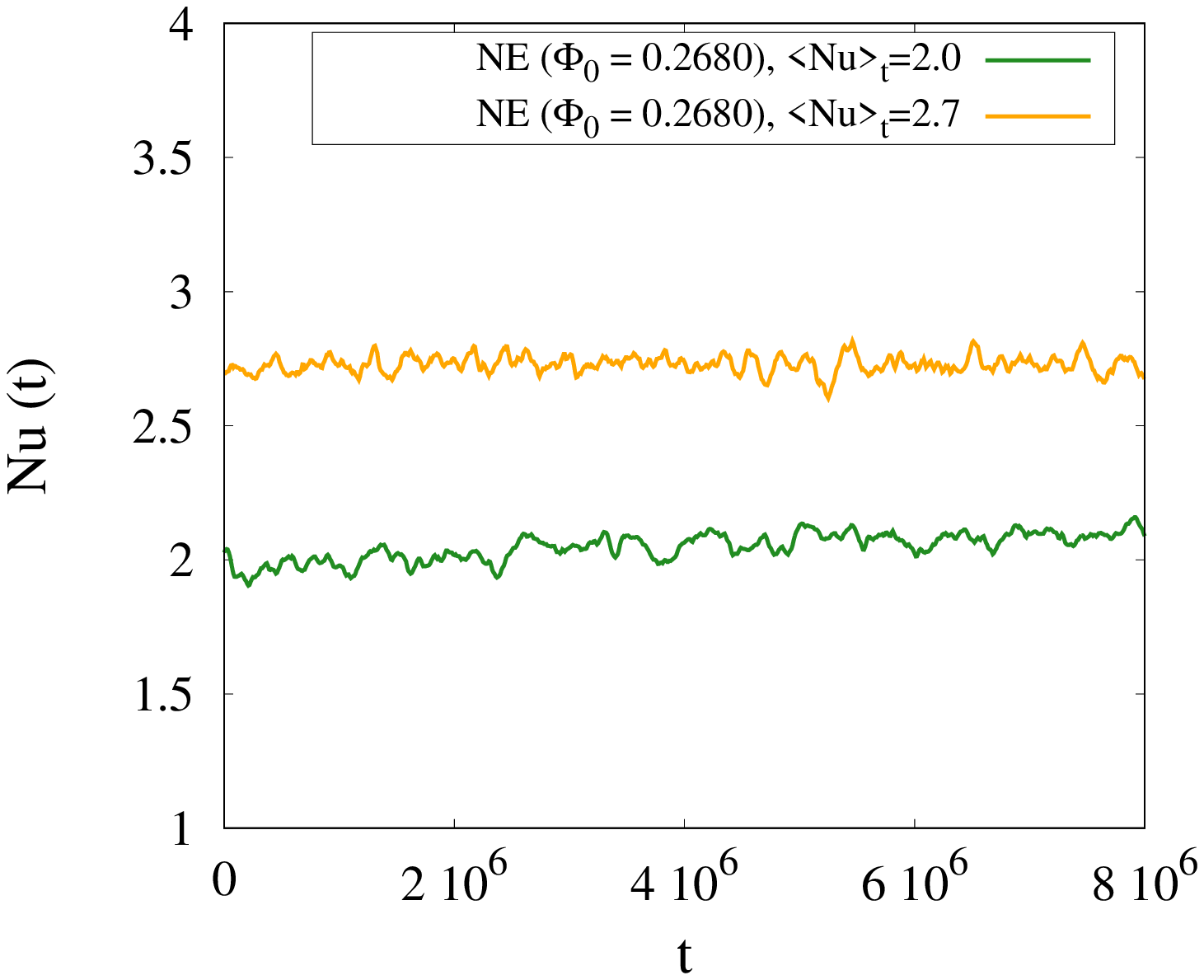} & \includegraphics[width=.315\linewidth]{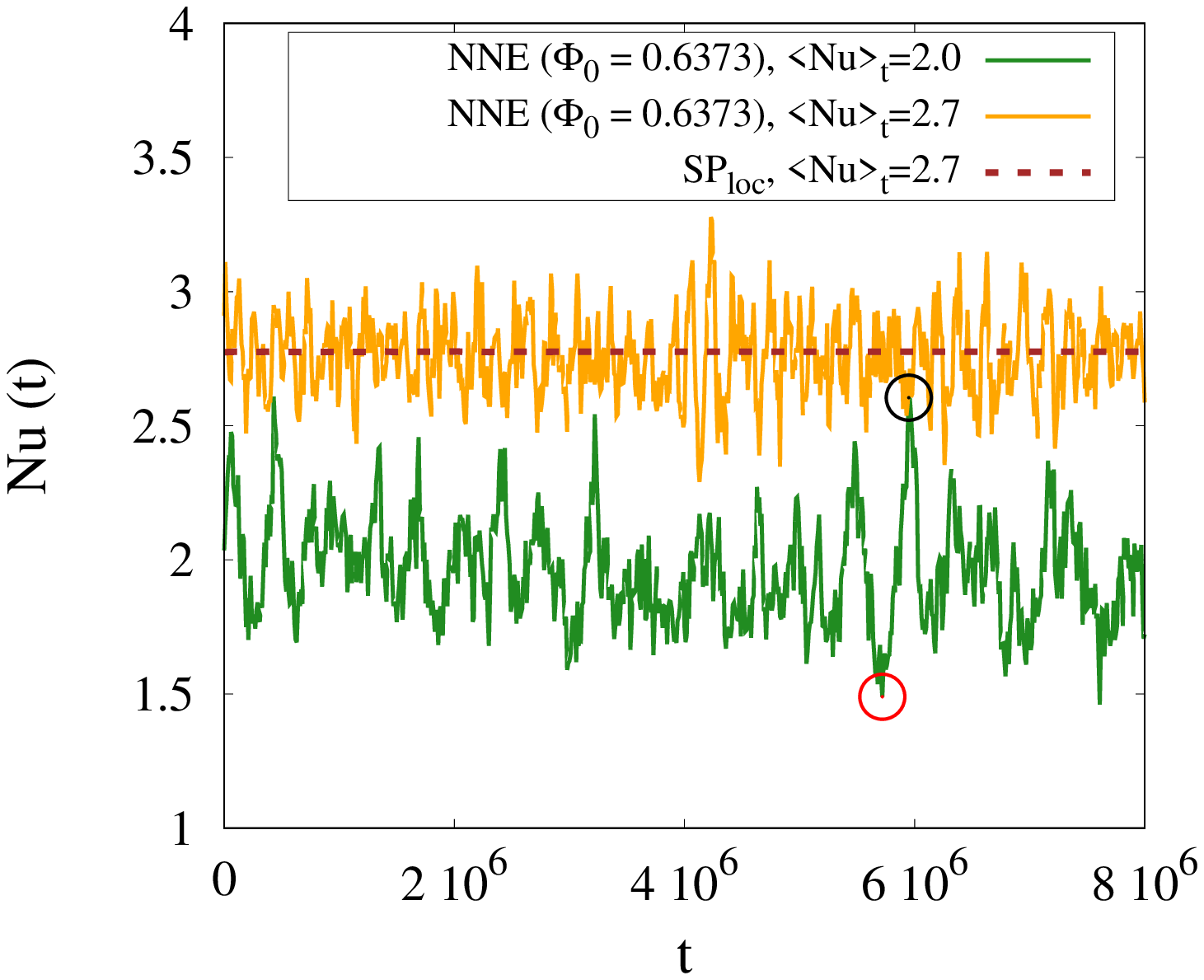} & \includegraphics[width=.315\linewidth]{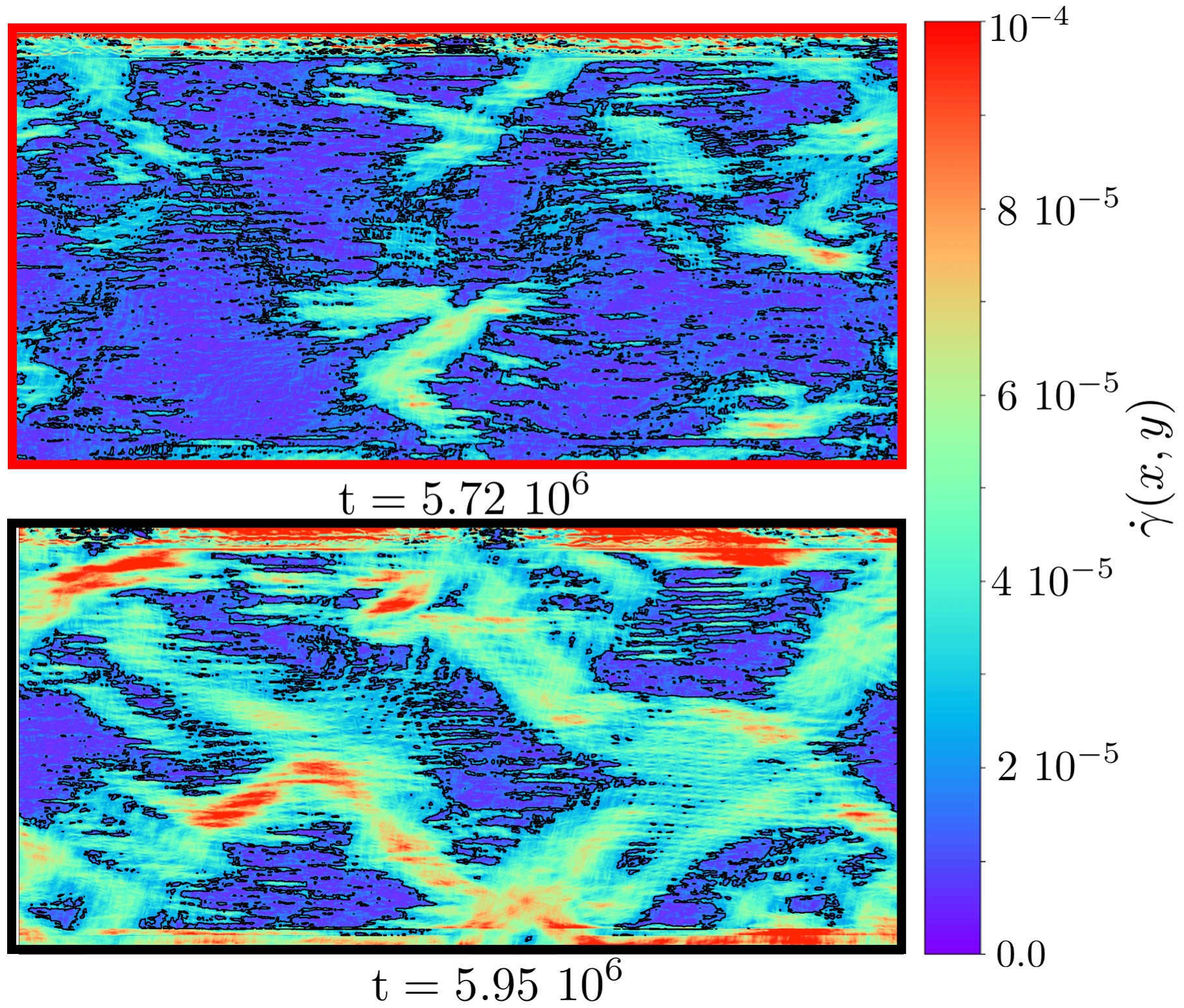} \\
\small (a) & \small (b) & \small      (c)\\
%\includegraphics[width=.3\linewidth]{fig2B.eps}\\
%\small (b) \\
%\includegraphics[width=.3\linewidth]{fig2C.png}\\
%\small      (c)
\end{tabular}
\caption{Time behaviour of the Nusselt number (cfr. Eq.~\eqref{eq:NuTime}) for both Newtonian (NE) and non-Newtonian (NNE) emulsions. Panel (a): we report $\mbox{Nu}$(t) as a function of time $t$ for a NE. The buoyancy forces are changed to fix the time-averaged Nusselt number $\langle \mbox{Nu}\rangle_t$ (cfr. Table \ref{table2}). Panel (b): same plot as panel (a) for the NNE. We also report the time behaviour for the $\mbox{Nu}$(t) for a single-phase ($\mbox{SP}_{\mbox{\tiny loc}}$) with local rheology obtained from the flow-curve of NNE (cfr. Fig. 3 in the ESI section). Concerning the NNE case, we emphasise that the same time-averaged Nusselt number $\langle \mbox{Nu} \rangle_t$ can be obtained from the $\mbox{SP}_{\mbox{\tiny loc}}$ system only by decreasing the buoyancy amplitude (details are given in Table~\ref{table2}). In the ESI section we include a simulation movie to highlight the different dynamics for NE and NNE. Panel (c): 2D-maps of the local shear $\dot{\gamma}(x,y)$ at selected times (circles in panel (b)) for the NNE. All dimensional quantities are reported in simulation units.}\label{fig:NuVsTime}
%\end{minipage}
\end{figure*}
%%%%%%%%%%%%%%%%%%%%%%%%%%%%%%%%%%%%%%%%%%%%%%%%%%%%

In doing so, a first marked difference emerges in the comparison between NE and NNE. While NE can flow with a Nusselt number that is only mildly dependent on time, NNE shows neat and larger fluctuations in the Nusselt number (cfr. Fig.~\ref{fig:NuVsTime}). To dig deeper into this phenomenology, we also report in Fig.~\ref{fig:NuVsTime}(c) the 2D-maps of the local shear $\dot{\gamma}$ in correspondence of a local maximum/minimum in $\mbox{Nu}(t)$ for NNE. These maps clearly show the coexistence of spatial regions, of different extent, at very small and larger shear rates $\dot{\gamma}$, respectively. In other words, while in correspondence of a maximum in the Nusselt number, the system is predominantly fluidised with a little number of small shear rates (i.e., large viscosity) regions, in correspondence of a minimum in the Nusselt number, the reversed situation holds.
We remark that for a homogeneous Newtonian fluid at these values of the Nusselt number the convective states are {\it time-independent}, hence it is natural to ask where these fluctuations come from. To get further insight into the problem, we studied also the case of a single-phase ($\mbox{SP}_{\mbox{\tiny loc}}$) fluid with a ``local'' closure for the effective dynamic viscosity. More precisely, we fitted the rheological curve of NNE displayed in Fig. 3 in the ESI section and extracted the effective ``local'' viscosity from the slope, $\eta_{\text{eff}}(\dot{\gamma})=d \Sigma(\dot{\gamma})/d \dot{\gamma}$; we then ran a numerical simulation with Eq.~\eqref{eq:NS} with the so constructed $\eta_{\text{eff}}(\dot{\gamma})$, however without droplets. In this way, we are simulating a homogeneous fluid inheriting the complex rheology of the emulsion via a local relationship between the dynamic viscosity and the local shear rate. Also in this case, for an optimal comparison, a different buoyancy has been imposed, such as to keep the time-averaged Nusselt number $\langle\mbox{Nu}\rangle_t$ fixed. Table~\ref{table2} shows the numerical values of the buoyancy amplitudes $\alpha g \Delta T$ used in the simulations. With respect to the NNE, we remark that it is necessary to reduce the buoyancy amplitude $\alpha g \Delta T$ of about $60\%$ in the $\mbox{SP}_{\mbox{\tiny loc}}$ case in order to obtain the same heat transfer of the NNE. In other words, the $\mbox{SP}_{\mbox{\tiny loc}}$ case does not reproduce the same time-averaged Nusselt number $\langle\mbox{Nu}\rangle_t$ for the same buoyancy amplitude of the NNE. Moreover, as shown in Fig.~\ref{fig:NuVsTime}(b), the fluid with the ``local'' rheology does not show any fluctuations of $\mbox{Nu}(t)$. The conclusion that we draw is that the non-Newtonian rheology is not sufficient to observe the {\it neat} fluctuations; rather, we must have non-Newtonian rheology supplemented with the presence of finite-size droplets. We deemed, therefore, appropriate to inspect this phenomenology from a Lagrangian viewpoint, that is looking at the relevant observables along a droplet trajectory; in particular, inspired by Lagrangian studies of turbulent RB convection~\cite{Gasteuil07,Schumacher09}, we focus on the droplet Nusselt number $\mbox{Nu}^{(\mbox{\tiny drop})}_i$. The definition of this Lagrangian observable is constructed in such a way that the global Nusselt number $\mbox{Nu}$ may be seen as the sum over the local contributions of the single droplets, i.e.,
\begin{equation}\label{eq:NuMatch}
\mbox{Nu}(t) = \frac{1}{N_{\mbox{\tiny droplets}}}\sum_{i=1}^{N_{\mbox{\tiny droplets}}} \mbox{Nu}^{(\mbox{\tiny drop})}_{i}(t). 
\end{equation}
A good candidate to satisfy Eq.~\eqref{eq:NuMatch} is the droplet Nusselt number defined as:
\begin{equation}\label{eq:NuDrops}
\mbox{Nu}^{(\mbox{\tiny drop})}_{i}(t) = \frac{u^{(i)}_y(t)  T^{(i)}(t) - \kappa (\partial_yT)^{(i)}(t)}{\kappa \frac{\Delta T}{H}}.
\end{equation}
where $u^{(i)}_y(t)=u_y(\mathbf{X}_i(t),t)$, $T^{(i)}(t)=T(\mathbf{X}_i(t),t)$ and $(\partial_yT)^{(i)}(t)=\partial_y T(\mathbf{X}_i(t),t)$ are the fluid velocity, temperature, and temperature gradient evaluated at the position of the $i$-th droplet centre-of-mass, $\mathbf{X}_i(t)$ ($i=1...N_{\mbox{\tiny droplets}}$), respectively.
%%%%%%%%%%%%%%%%%%%%%%%%%%%%%%%%%%%%%%%%%%%%%
\begin{figure}[t!]
\centering
\begin{tabular}{c}
\includegraphics[width=.99\linewidth]{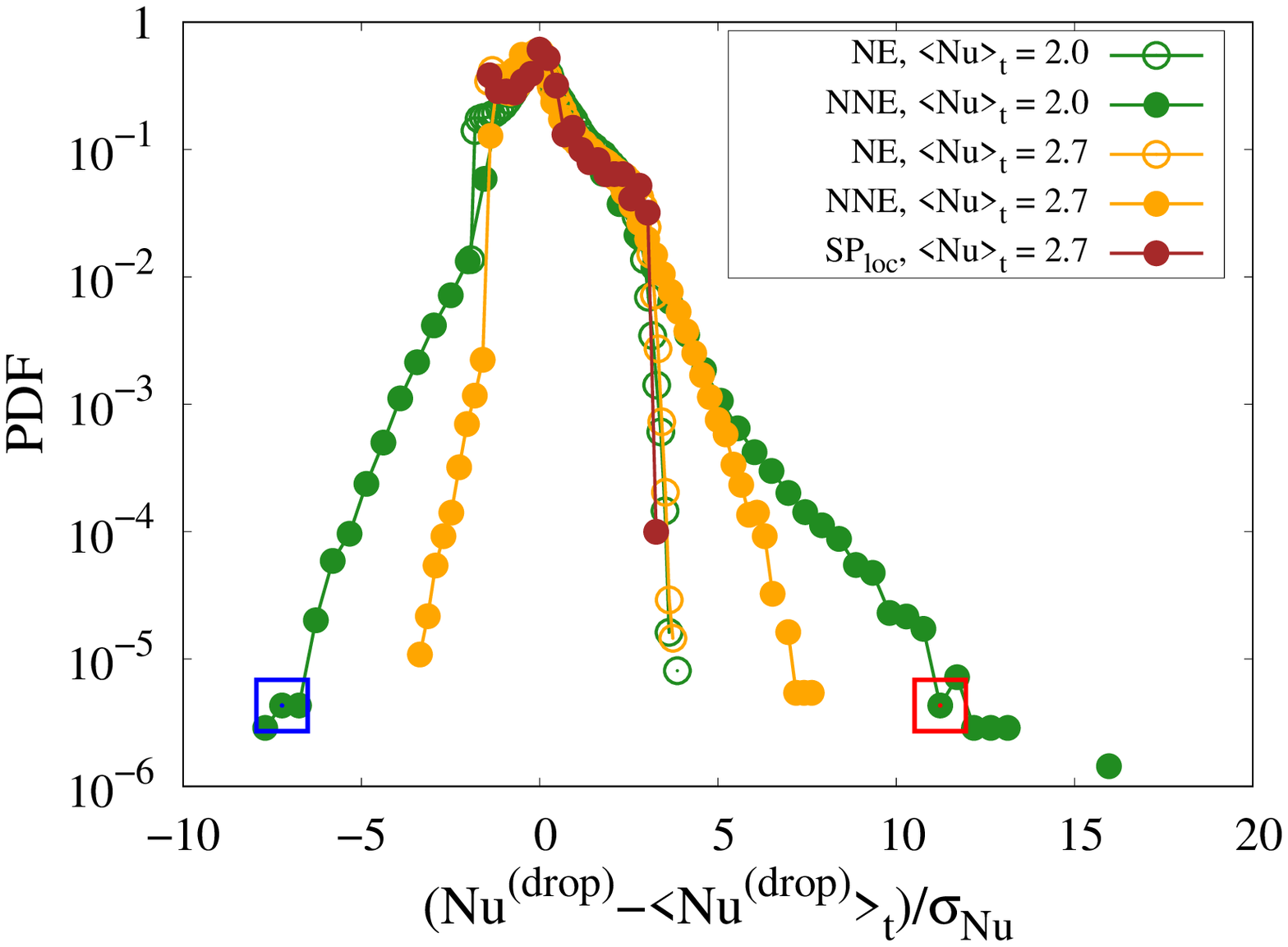}\\
\small (a) \\
\includegraphics[width=.98\linewidth]{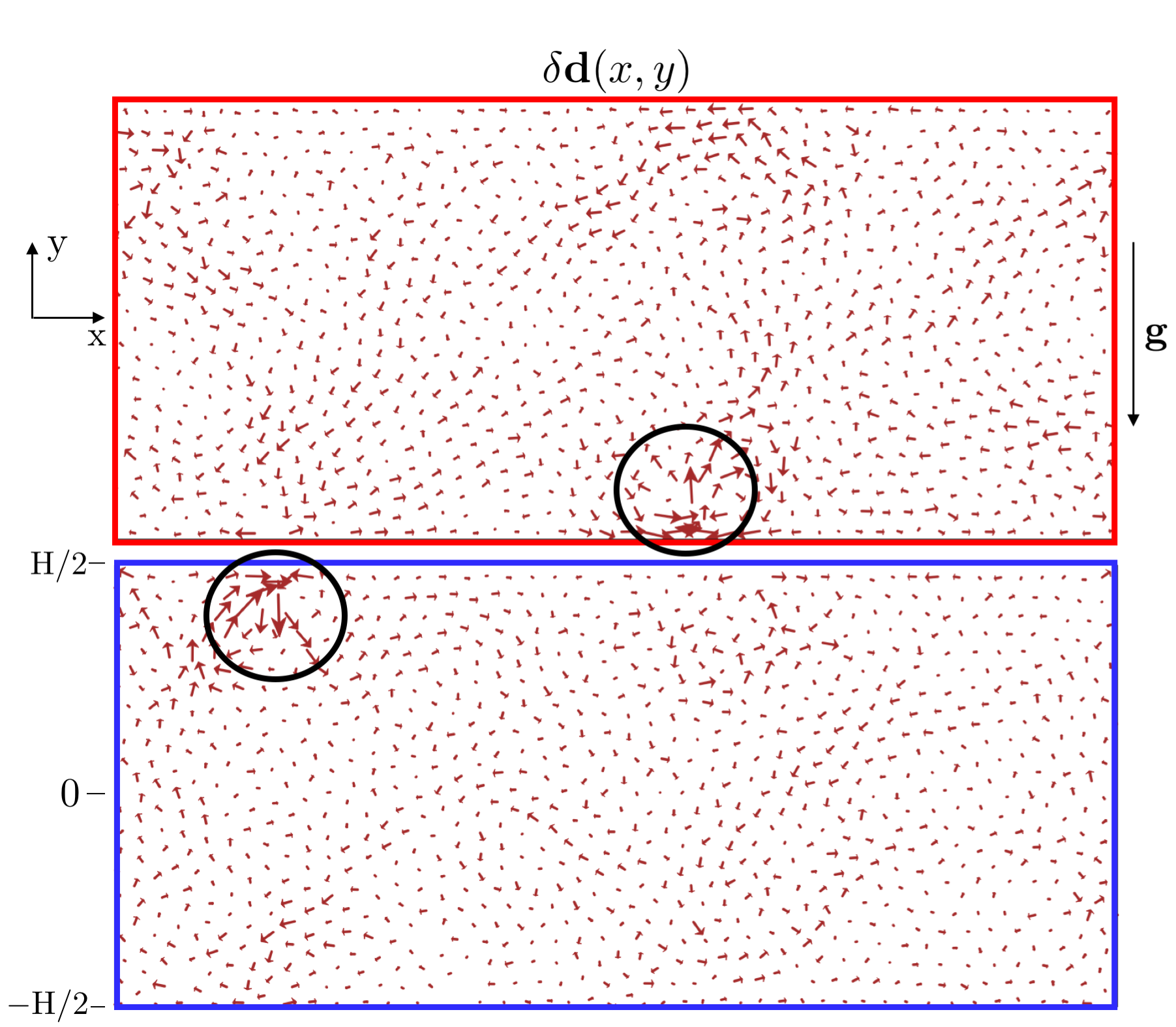}\\
\small (b) \\
\end{tabular}
\caption{Panel (a): Log-Lin probability distribution function (PDF) of the droplet Nusselt number (cfr. Eq.~\eqref{eq:NuDrops}) for simulations reported in Fig.~\ref{fig:NuVsTime} (see text for details). To obtain the statistics of the droplet Nusselt number, we consider the observable~\eqref{eq:NuDrops} and we analyse its statistical properties for all the droplets and for all times. It is shown in units of the standard deviation with respect to the average value $\langle \mbox{Nu}^{(\mbox{\tiny drop})} \rangle_{t}$. Panel (b): we report snapshots of vectorial displacement fluctuations (cfr. Eq.~\eqref{eq:displacementU}), where it is possible to observe those events that contribute to positive (top red panel) and negative (bottom blue panel) tails of the PDF of the NNE. \label{fig:PDFNu}}
\end{figure}
%%%%%%%%%%%%%%%%%%%%%%%%%%%%%%%%%%%%%%%%%%%%%
In Fig.~\ref{fig:PDFNu}(a) we report the PDF of the droplet Nusselt number for the numerical simulations previously analysed in Fig.~\ref{fig:NuVsTime}. Here we extract the PDF by analysing the Nusselt number of all droplets at all times. For the sake of comparison, we show data with the $x$-axis given in units of the standard deviation with respect to the average value $\langle \mbox{Nu}^{(\mbox{\tiny drop})} \rangle_{t}$~\footnote{The average value $\langle \mbox{Nu}^{(\mbox{\tiny drop})} \rangle_{t}$ is computed by considering all droplets at all times.}.
Being the $\mbox{SP}_{\mbox{\tiny loc}}$ simulation without droplets, for the computation of $\mbox{Nu}^{(\mbox{\tiny drop})}_{i}$ we took an Eulerian viewpoint and divided the computational domain in boxes (in number equal to $N_{\mbox{\tiny droplets}}$ of NNE case) and computed $\mbox{Nu}^{(\mbox{\tiny drop})}_{i}(t)$ for each box. The most evident result regards the PDFs tails, shown in Fig.~\ref{fig:PDFNu}(a): while the PDF for the NE drops to zero at roughly 5-6 standard deviations, the PDF for the NNE exhibits more pronounced tails, up to 10-15 standard deviations. We have analysed the events contributing to such fat tails and found that, in correspondence of the ``extreme'' events (either in the positive or the negative tail), neat vectorial displacement fluctuations $\delta {\bf d}$ (cfr. Eq.~\eqref{eq:displacementU}) are observed. These fluctuations are nothing but droplets rearrangements which contribute to ``boost'' the thermal convection, hence providing enhanced positive tails in the PDF of the droplet Nusselt number (red box in the top panel of Fig.~\ref{fig:PDFNu}(b)); rearrangements may also inhibit convective transport, hence a contribution to the negative tail of the PDF (blue box in the bottom panel of Fig.~\ref{fig:PDFNu}(b)). Notice also that such ``extreme'' events are located within the boundary layers. We remark that the enhancement of the tails appears only in the presence of finite-size droplets, whereas the $\mbox{SP}_{\mbox{\tiny loc}}$ model does not show such pronounced tails, being closer to the NE case.
%\begin{figure}[t!]
%\centering
%\includegraphics[width=0.95\linewidth]{fig3A.eps}\\
%\vspace{0.3cm}
%\hspace{0.2cm}
%\includegraphics[width=.92\linewidth]{fig3B.png}
%\caption{Top panel: Log-Lin probability distribution function (PDF) of the droplet Nusselt $\mbox{Nu}_{i}$ (cfr. Eq.~\eqref{eq:NuDrops}) for data reported in Fig.~\ref{fig:NuVsTime}. Data are reported in standardized form, based on the average $\langle \mbox{Nu}_{i} \rangle$ and standard deviation $\sigma$. Bottom panel: we report vectorial displacement fluctuations (cfr. Eq.~\ref{eq:displacementU}) where it is possible to observe those events that contribute to positive and negative tails of the PDF of the CE. \label{fig:PDFNu}}
%\end{figure}
%%%%%%%%%%%%%%%%%%%%%%%%%%%%%%%%%%%%%%%%%%%%%
The analysis performed in Fig.~\ref{fig:PDFNu} helps in further elucidating the large scale fluctuations in the Nusselt number observed in Fig.~\ref{fig:NuVsTime}. In particular, it gives some hints on the physical mechanism that allows the system to display the switch shown in Fig.~\ref{fig:NuVsTime}(c). If the system is almost entirely non-fluidised, it can change to a situation where it is predominantly fluidised if non-local correlations are active in the system. The same holds for a system that is predominantly fluidised and switches back to an almost entirely non-fluidised state. Spatially extended correlated zones are also expected by looking at the maps of $\delta {\bf d}(x,y,t)$ reported in Fig.~\ref{fig:PDFNu}(b), where collective ``bursts'' of $\delta {\bf d}(x,y,t)$ appear. In fact, in the absence of space correlations, ``bursts'' of activity would be unable to propagate in the system and trigger the switching of a substantially large part of the system in another state. These facts said it comes as a logical consequence to study the observable $\delta {\bf d}(x,y,t)$ to better corroborate the existence of non-trivial correlations in the system. To this aim, we first average the field $\delta {\bf d}(x,y,t)$ in the $x$-direction, i.e., $\tilde{\delta {\bf d}}(y,t) = \langle \delta {\bf d} (x,y,t)\rangle_x$.
%%%%%%%%%%%%%%%%%%%%%%%%%%%%%%%%%%%%%%%%%%%%%
\begin{figure}[t!]
\centering
\begin{tabular}{c}
\includegraphics[width=.99\linewidth]{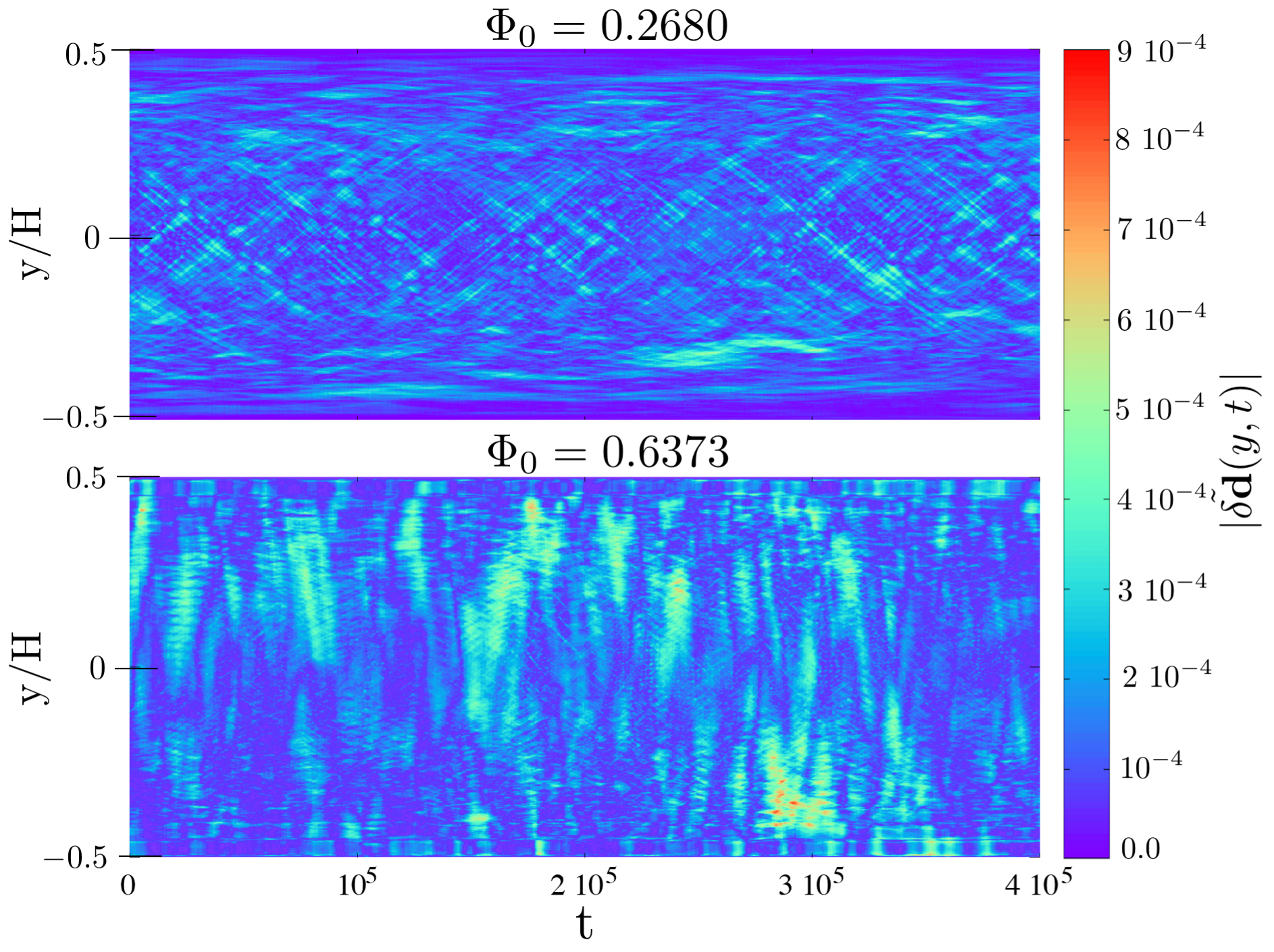}\\
\small (a) \\
\includegraphics[width=.99\linewidth]{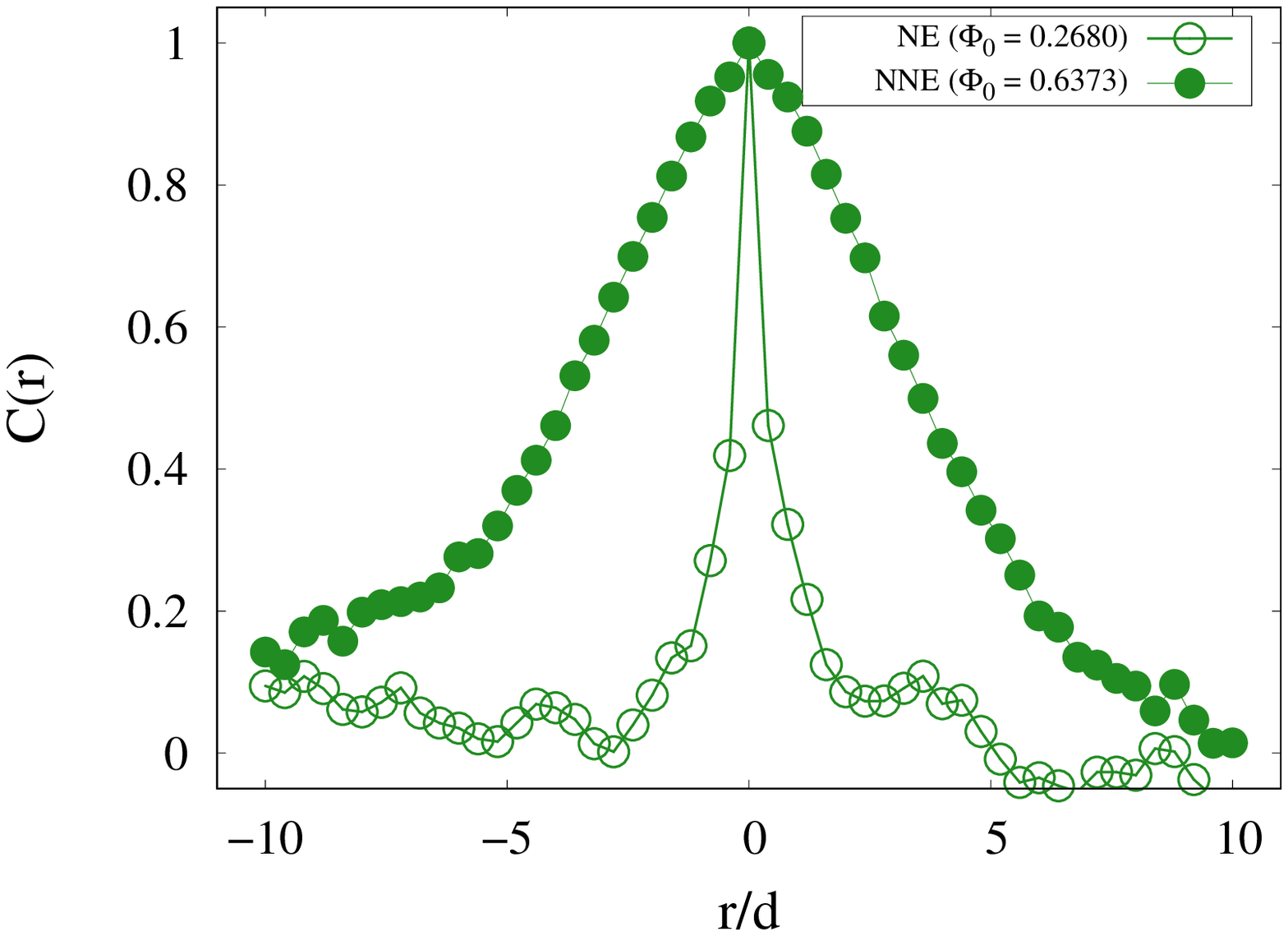}\\
\small (b) \\
\end{tabular}
\caption[figure]{Panel (a): space-time evolution of the absolute value of the $x$-averaged displacement fluctuations, $\tilde{\delta {\bf d}}(y,t) = \langle \delta {\bf d} (x,y,t)\rangle_x$ (see text for details) in the $(y,t)$ plane. Panel (b): correlation function $\mbox{C}(r)$ of the absolute value of $\tilde{\delta{\bf d}}(y,t)$ (cfr. Eq.~\eqref{eq:correlation}) for both NE and NNE. The variable $\mbox{r}$ is normalised with the mean droplet diameter $d$. All dimensional quantities are reported in simulation units.}\label{fig:correlation}
\end{figure}
%%%%%%%%%%%%%%%%%%%%%%%%%%%%%%%%%%%%%%%%%%%%%
The space-time evolution of the displacement fluctuations is reported in Fig.~\ref{fig:correlation}(a), where we plot the absolute value of $\tilde{\delta {\bf d}}(y,t)$ in the $(y,t)$ plane. It is seen that for the NNE the displacement fluctuations depart from zero coherently in extended space regions, predominantly close to the boundaries. Such space coherence persists for some finite time. This is in marked contrast with the observations for the NE, where the space-time coherence is visibly lost. Finally, to unveil more quantitatively the difference in space correlations between NE and NNE, we have computed the spatial correlation function $\mbox{C}(r)$. To this aim, we have adapted the definitions of previous literature studies~\cite{Lancaster97,Berthier11,Benzietal14} to the absolute value of $\tilde{\delta {\bf d}}(y,t)$:
\begin{equation}\label{eq:correlation}
\mbox{C}(r) =\dfrac{\langle |\tilde{\delta {\bf d}}(0,t)| \ | \tilde{\delta {\bf d}}(r,t)|\rangle_t- \langle|\tilde{\delta {\bf d}}(0,t) | \rangle_t \ \langle |\tilde{\delta {\bf d}}(r,t)| \rangle_t}{\sigma(0) \sigma(r)}
\end{equation}
where $-H/2 < r < +H/2$ and $\sigma(0)$($\sigma(r)$) is the standard deviation of $|\tilde{\delta {\bf d}}(0,t)|$ ($|\tilde{\delta {\bf d}}(r,t)|$). In Fig.~\ref{fig:correlation}(b) we show $\mbox{C}(r)$ for both NE and NNE with the $x$-axis normalised by the mean droplet diameter $d$: while for the NE case the correlation rapidly decays to zero within a distance of the order of single droplet diameter, the NNE emulsion shows larger correlation extending in space for a markedly larger distance. \\
It is worth noting that two reference cases studied here are representative of two different ``categories'' of emulsions: a dilute, Newtonian, emulsion and a concentrated emulsion. A continuous scan of the volume fraction may well reveal intermediate situations, with an incipient non-Newtonian character (manifesting itself, e.g., in the form of a weak shear-thinning), whereby the observed phenomenology -- i.e., non-Gaussian temporal statistics of the heat flux and enhanced space correlation -- falls somehow in between the two instances considered here. 

%%%%%%%%%%%%%%%%%%%%%%%%%%%%%%%%%%%%%%%%%%%%
\section{Conclusions}\label{sec:conclusions}
%%%%%%%%%%%%%%%%%%%%%%%%%%%%%%%%%%%%%%%%%%%%%

We analysed the heat transfer properties of a model emulsion in the Rayleigh-B\'{e}nard (RB) set-up, where the emulsion is placed in a confined cell between two parallel walls at different temperatures (a hot bottom and a cold top wall). The droplet concentrations $\Phi_0$ in the emulsions have been chosen to range systematically from very dilute cases (Newtonian emulsions) to situations with larger concentrations, where the emulsion behaves as a non-Newtonian fluid. We explored the heat transfer properties while keeping the droplet size finite, thus disclosing insights into the way a continuum picture (i.e., point-like droplets) is changed by the finite-size effects induced by a non-zero extension of the droplets. It is well known that the transition to convection of a homogeneous Newtonian system is accompanied by the onset of steady flow and time-independent heat flux; in marked contrast, the heterogeneity of emulsions brings in an additional and previously unexplored phenomenology. We find that the heat transport efficiency (i.e., the Nusselt number, $Nu$) displays a non-stationary character in time at fixed buoyancy intensity: while its time-average decreases at increasing $\Phi_0$, the fluctuations around the mean value increase. Besides, due to the convective dynamics, the emulsion develops a non-homogeneous droplet distribution across the cell.\\
In the attempt of capturing the time-averaged Nusselt number $\langle \mbox{Nu}\rangle_t$ at changing droplet concentration $\Phi_0$, we pursued the idea of considering a single-phase (SP) system, equipped with a suitable choice of viscosity $\eta^{\mbox{\tiny SP}}$ that allows the SP system to display the same heat transport efficiency of the emulsions. Starting from the knowledge of the shear rheology for the emulsions $\eta_{\text{eff}}(\Phi_0)$, we investigated the suitable protocol that allows constructing $\eta^{\mbox{\tiny SP}}$. Due to the non-homogeneous droplet distribution across the cell, we have explored the possibility that $\eta^{\mbox{\tiny SP}}$ could acquire a space-dependence. A quantitative analysis reveals that this local viscosity must be properly supplemented with a spatial averaging procedure (``coarse-graining''), over a scale that is of the order of the droplet size. In this part, we deliberately discussed results on emulsion concentrations resulting in Newtonian responses. Indeed, further increasing the droplet concentration would produce a non-Newtonian emulsion with an effective viscosity that depends on the shear-rate. Any kind of attempt that we tried to capture the time-averaged Nusselt number for such non-Newtonian emulsions failed. These findings raise interesting questions as to the precise meaning of viscosity when the assumption of continuity breaks down and scales involved become of the order of the size of constituents. It is known from the literature on the rheology of highly concentrated emulsions~\cite{Goyon08,Goyon10,BenziSbragaglia16,Derzsi17} that non-local effects are present at such small scales. Non-local effects impact significantly the flow properties and they can be reabsorbed into a continuum formulation by introducing an effective diffusivity in the dynamical equations for the ``fluidity'' field (i.e., inverse viscosity). If and how this is possible for the convective systems studied in this paper, certainly deserves future scrutiny. \\
Heat-flux fluctuations have been studied via a systematic comparison on the heat transfer properties between two representative emulsion concentrations: a Newtonian emulsion (NE) exhibiting Newtonian rheology, and a non-Newtonian emulsion (NNE) exhibiting shear-thinning rheology with a marked increase of the viscosity at low shear rates. We have shown that the presence of non-Newtonian rheology and finite-size droplets conspire to trigger the emergence of neat fluctuations in the Nusselt number, corresponding to the switching between two qualitatively different system configurations, with a predominance of fluidised (i.e., a maximum of the Nusselt number) and non-fluidised (i.e., a minimum of the Nusselt number) regions. This goes together with the emergence of fat tails in the statistics of the local Nusselt number, i.e., the Nusselt number at the droplet scale. Overall, the convective phenomenology for the NNE is attributed to the emergence of a finite correlation between distant droplets, which we have unveiled via the analysis of the displacement fields. The correct way to capture these temporal fluctuations is not clear at this stage: they can be measured and characterised in the simulations with the emulsions, but the specific way to embed them in a continuum approach warrants a dedicated study. \\
Overall, all our findings suggest that any approach aiming at a quantitative description of heat transfer in fluid-fluid dispersions at scales comparable to the size of the constituents must take into account the discrete nature of such complex fluids.

\section*{Acknowledgements}

Funding support to lead these results was received from the European Research Council under the Horizon 2020 Programme Grant Agreement n. 739964 ("COPMAT"). The authors acknowledge Giacomo Falcucci, Fabio Bonaccorso for useful support. Some of the simulations were performed on Jeeg, the graphical-accelerated facility of the University of Naples ``Parthenope''. Jeeg was acquired with the Italian Government Grant PAC01\_00119 MITO ``Informazioni Multimediali per Oggetti Territoriali'', with Elio Jannelli as the principal investigator.

%\scriptsize{
\bibliographystyle{rsc}
\bibliography{francesca}

\end{document}